\documentclass[aps,prl,amsmath,amssymb,floatfix,twocolumn,amsmath,superscriptaddress,twocolumn,nofootinbib,tighten,letterpaper]{revtex4-2}
\usepackage[colorlinks,linkcolor=blue,citecolor=blue,urlcolor=blue]{hyperref}
\usepackage{multirow}
\usepackage{subfigure}
\usepackage{color}
\usepackage{mathrsfs}
\usepackage{hyperref}
\usepackage[normalem]{ulem}
\usepackage{bm}

\usepackage{amssymb}   
\usepackage{amsmath}
\renewcommand\vec[1]{\ensuremath\boldsymbol{#1}} 

\usepackage{amsfonts, relsize, color}
\usepackage{graphics}
\usepackage{graphicx}
\usepackage{subfigure}
\usepackage{hyperref}
\usepackage{color}
\usepackage{comment}

\begin{document}

\title{From local spin nematicity to altermagnets: Footprints of band topology}

\author{Sanjib Kumar Das}
\affiliation{Department of Physics, Lehigh University, Bethlehem, Pennsylvania, 18015, USA}

\author{Bitan Roy}~\thanks{Corresponding author: bitan.roy@lehigh.edu}
\affiliation{Department of Physics, Lehigh University, Bethlehem, Pennsylvania, 18015, USA}

\date{\today}

\begin{abstract}
Altermagnets are crystallographic rotational symmetry breaking spin-ordered states, possessing a net zero magnetization despite manifesting Kramer's non-degenerate bands. Here, we show that momentum-independent local spin nematic orders in monolayer, Bernal bilayer, and rhombohedral trilayer graphene give rise to $p$-wave, $d$-wave, and $f$-wave altermagnets, respectively, thereby inheriting the topology of linear, quadratic and cubic free fermion band dispersions that are also described in terms of angular momentum $\ell=1,\; 2$, and $3$ harmonics in the reciprocal space. The same conclusions also hold inside a spin-triplet nematic superconductor, featuring Majorana altermagnets. Altogether, these findings highlight the importance of electronic band structure in identifying such exotic magnetic orders
in quantum materials. We depict the effects of in-plane magnetic fields on altermagnets, and propose spin-disordered alter-valley magnets in these systems.    
\end{abstract}

\maketitle

\emph{Introduction}.~Magnetic materials commonly appear inside modern-day electronic devices. When doped, often they also source  unconventional and high-temperature superconductors. Therefore, identifying new magnetic phases and materials are of both fundamental and technological importance, possibly paving a path toward the long sought room temperature superconductors.

Typically, magnetic materials are grouped into two families, ferromagnet and anti-ferromagnet. The former breaks only the time-reversal symmetry, thereby lifting the Kramer's degeneracy of electronic bands. It possesses a finite magnetic moment, resulting from a population imbalance between electrons with opposite spins. By contrast, the Kramer's degeneracy of electronic states is protected in an anti-ferromagnet, stemming from the simultaneous lifting of the time-reversal and inversion symmetries, yielding a net zero magnetization.

Recently, a new type of magnetic order has been proposed theoretically~\cite{Hayami2019, Ahn2019, Smejkal2020, Chen2020, JunweiLiu2021, Smejkal2022I, Smejkal2022II, Smejkal2022III, Turek2022, Libor2023, Mazin2023, Fakhredine2023, Gao2023, Run2023, Chi2023, Ouassou2023, Steward2023, Fernandes2024, Agterberg2024, Parshukov2024}, and unearthed in quantum materials~\cite{Feng2022, Reimers2023, Lovesey2023, Igor2023, Fedchenko2023, Bai2023, Grzybowski2023, Kluczyk2023, Takuya2023, Cuono2023, Lin2024, Lee2024, Guo2023}: \emph{altermagnets}. Despite lifting the Kramer's degeneracy, they manifest no net magnetization, a peculiarity accomplished at the cost of discrete crystallographic rotational symmetry with opposite signs for complementary spin projections. They are represented in terms of spherical harmonics ($Y^{m}_{\ell}$), taking a generic form ${\boldsymbol \sigma} Y_{\ell}^{m}(\theta, \phi) |\vec{k}|^{\ell}$. Vector Pauli matrix ${\boldsymbol \sigma}$ operates on the spin space, $m=-\ell, \cdots, \ell$, $\theta$ ($\phi$) is the polar (azimuthal) angle in the reciprocal space, and $\vec{k}$ is the momentum. This classification allows $p$-wave ($\ell =1$), $d$-wave ($\ell=2$), and $f$-wave ($\ell=3$) altermagnets, to name a few.

Although strongly correlated materials can in principle harbor such exotic magnetic orders, their non-locality or momentum-dependence can be energetically expensive, forcing us to the raise the following question. \emph{Can altermagnets emerge from momentum-independent local magnetic orders?} We show that its affirmative answer establishes a topology-based guiding principle to identify quantum materials, capable of fostering altermagnets.

This question arises from a seemingly unrelated topic, topological superconductors (TSCs), worth mentioning despite a short detour. Consider their prime member, the B-phase of $^3$He, a fully gapped $p$-wave paired state~\cite{Volovik2009}. It can emerge from \emph{local} or \emph{on-site} odd-parity Cooper pairing in three-dimensional Dirac materials~\cite{FuBerg2010, LiangFu2014}, also modeled in terms of odd-parity $p$-wave harmonics. Therefore, neutral Bogoliubov-de Gennes (BdG) quasiparticles inherit topology from normal state charged Dirac quasiparticles. Moreover, when such a local odd-parity pairing is projected on the Fermi surface, realized by intercalating or doping topological insulators, it takes the form of the B phase of $^3$He~\cite{Alavirad2017, RoyFosterNevidomskyy2019}. This one-to-one correspondence between the normal state band topology and paired state emergent topology guides us to identify candidate materials, fostering charged TSCs, with Cu$_x$Bi$_2$Se$_3$ and In$_x$Sn$_{1-x}$Te standing as promising candidates~\cite{Ando2011PRL, Ando2012PRL, Ando2013PRB}. A similar avenue has also been built to identify candidate materials for higher-order TSCs~\cite{Ghorashi2019, BRoy2020, RoyJuricic2021}. In light of these observations, the question from the last paragraph can be rephrased in the following way. \emph{How does electronic band topology get imprinted on altermagnets?}

\begin{figure}[t!]
\includegraphics[width=1.00\linewidth]{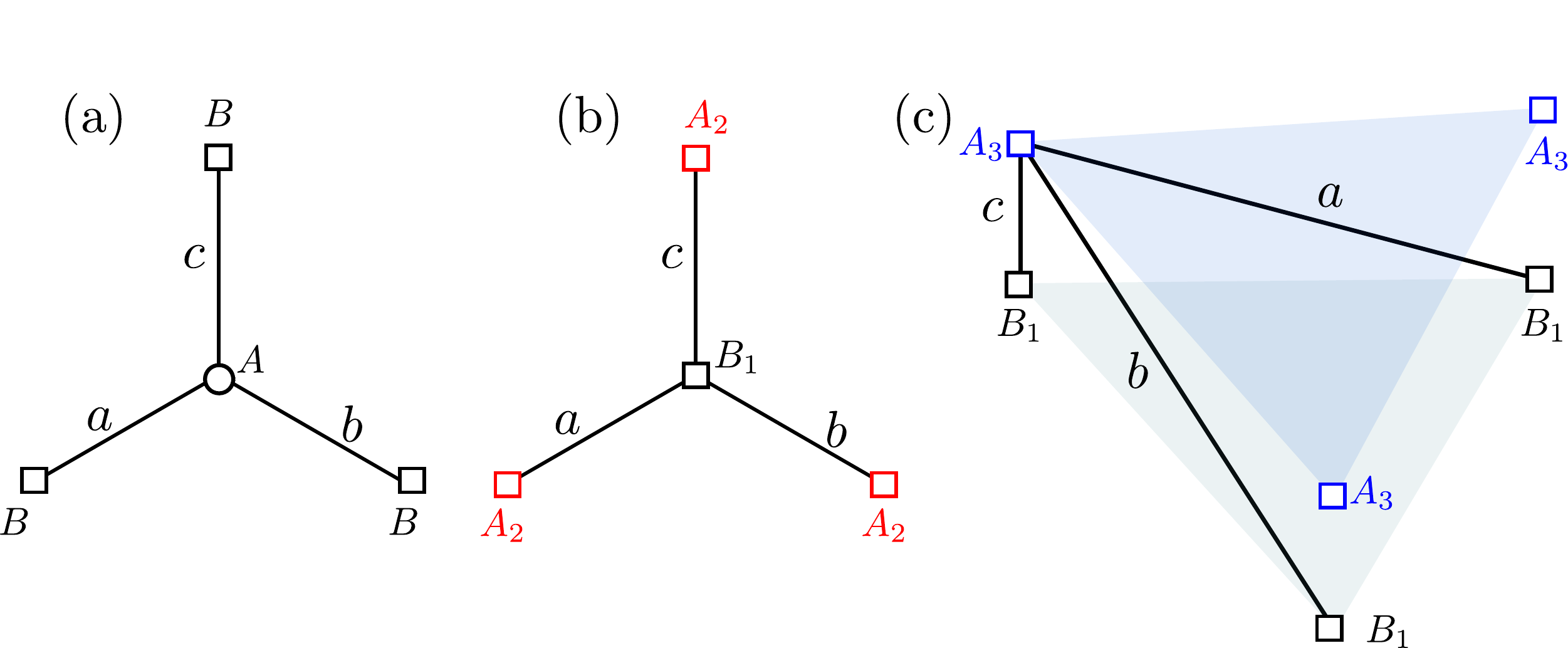}
\caption{Nematicity in (a) MLG, (b) BBLG, and (c) RTLG, resulting from hopping modulations between nearest-neighbor [(a)] or low-energy [(b) and (c)] sites, denoted by $a$, $b$, and $c$ with $|a| \neq |b| \neq |c|$. For $E_g$ ($E_u$) charge [spin] nematic orders $a$, $b$, and $c$ are purely real (imaginary) and bear the same [opposite] sign for opposite spin projections. In the spin-triplet $E_u$ nematic paired state, Cooper pair amplitudes $a$, $b$, and $c$ are purely imaginary with opposite signs for opposite spin projections. Here, $A$ and $B$ are two sublattices, and subscripts denote the layer index in (b) and (c)~\cite{SMaltermagnet}.    
}~\label{fig:Fig2}
\end{figure}

\begin{figure}[t!]
\includegraphics[width=1.00\linewidth]{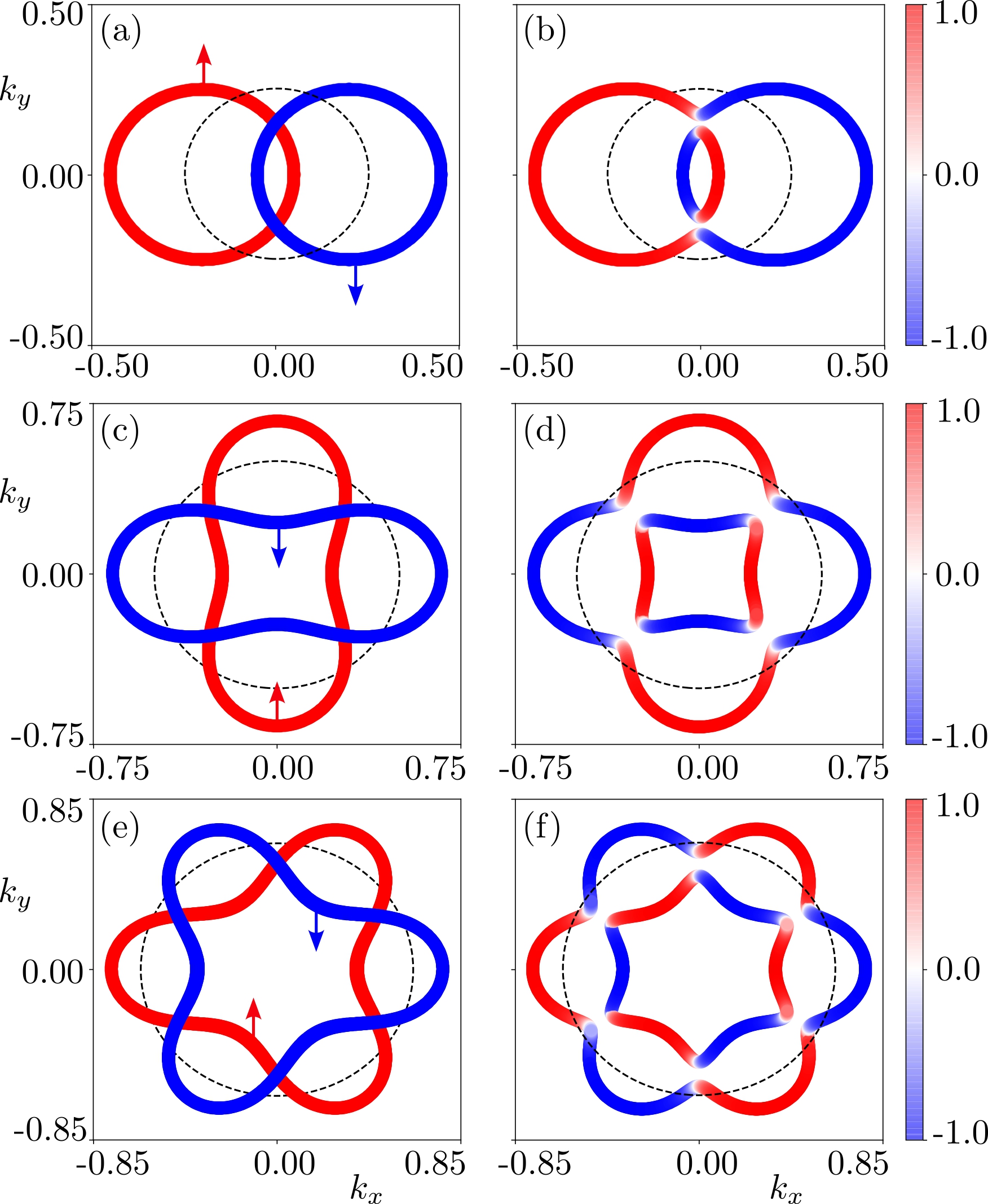}
\caption{Constant energy ($E=0.25$) contours, yielding Fermi surfaces at chemical doping $\mu=E=0.25$, near the $+{\bf K}$ valley for spin up ($\uparrow$) and down ($\downarrow$) electrons in the presence of local spin nematic orders (belonging to the $E_g$ or $E_u$ representation) without the Zeeman coupling in (a) MLG, (c) BBLG and (e) RTLG, displaying $p$-, $d$-, and $f$-wave altermagnets, respectively. We set $\Delta_j=0.2$ and $\theta_j=0$, where $j=E_g$ and $E_u$ [Eq.~\eqref{eq:spinnematic}]. The numbers in the color bar represent the spin projection in the $z$-direction (in units of $\hbar/2$). It is zero where the contours for opposite spin projections cross. They get split by a Zeeman coupling ($\Delta_Z$) of an in-plane magnetic field, as shown for (b) MLG ($\Delta_Z=0.025$), (d) BBLG ($\Delta_Z=0.05$) and (f) RTLG ($\Delta_Z=0.05$). Near the $-{\bf K}$ valley, the spin projection on each contour gets reversed (stays the same) in MLG and RTLG (BBLG) for the $E_g$ altermagnet. For the $E_u$ altermagnet, this correspondence is exactly the opposite. Momentum $\vec{k}$ is measured about the valley momentum ${\bf K}$. We set $\alpha_\ell=1$ [Eq.~\eqref{eq:freeHamil}]. Black dashed lines represent spin degenerate Fermi surface of free fermions. For each spin projection, the Fermi momentum determining the amplitude of the spin-resolved longitudinal conductivity (electrical or thermal) along any direction is the distance between the origin of the constant energy contour to a point on it in that direction.    
}~\label{fig:Fig1}
\end{figure}

Such broadly defined questions can be efficiently answered by considering minimal model Hamiltonian for crystalline graphene heterostructures. Here, we focus on monolayer graphene (MLG), Bernal bilayer graphene (BBLG), and rhombohedral trilayer graphene (RTLG) displaying linear, bi-quadratic, and bi-cubic touching of valence and conduction bands at two inequivalent corners of the hexagonal Brillouin zone, described by $p$-wave, $d$-wave, and $f$-wave harmonics in the momentum space, respectively~\cite{GrapheneRMP2009}, see Eq.~\eqref{eq:freeHamil}. In such systems, we show that local spin-nematic orders, transforming under the irreducible $E_g$ or $E_u$ representation of the $D_{3d}$ group, see Eq.~\eqref{eq:spinnematic} and Fig.~\ref{fig:Fig2}, give birth to emergent $p$-wave, $d$-wave, and $f$-wave altermagnets, respectively, inheriting topology from the normal state band dispersion, as shown by diagonalizing the effective single-particle Hamiltonian in the ordered phases [Eq.~\eqref{eq:altmagHamil}]. Results are depicted in Fig.~\ref{fig:Fig1}. By the same token, a spin-triplet nematic superconductor, belonging to the $E_u$ representation, fosters altermagnet for neutral Majorana fermions, hereafter coined Majorana altermagnet. We recognize that the valley or pseudo-spin degree of freedom permits a spin-disordered charge nematic order, leading to (un-nested) displaced or distorted Fermi surfaces near two valleys, a phase hereafter named alter-valley magnet. Although pristine MLG does not exhibit any symmetry breaking (due to vanishing density of states therein), we investigate its altermagnetic properties as a preparatory step toward studying time reversal symmetry breaking spin-nematic phases in BBLG and RTLG, where the corresponding ordering tendencies become prominent due to the constant and diverging density of states therein, respectively.

\emph{Free fermions}.~The continuum models, resulting from a minimal tight-binding Hamiltonian involving nearest-neighbor intra-layer ($t$) and inter-layer dimer ($t_\perp$) hopping~\cite{GrapheneRMP2009}, in MLG, BBLG and RTLG graphene, detailed in Sec.~S1 of the Supplemental Material (SM)~\cite{SMaltermagnet}, in a 16-component Nambu-doubled spinor basis read as   
\allowdisplaybreaks[4]
\begin{equation}~\label{eq:freeHamil}
\hat{h}_{\ell}(\vec{k}) = \alpha_\ell |\vec{k}|^\ell \left[ \Gamma^{1}_\ell \cos(\ell \phi) - \Gamma^{2}_\ell \sin(\ell \phi) \right] 
-\Delta_{\rm Z} \Gamma_{0100}, 
\end{equation}   
with $\ell=1,2$, and $3$, respectively. Here, $\alpha_\ell \sim (ta)^\ell/t^{\ell-1}_\perp$, bearing the dimension of Fermi velocity (inverse mass) for $\ell=1$ (2), for example, $a$ is the lattice spacing, $\Gamma^{1}_{1/3}=\Gamma_{3031}$, $\Gamma^{2}_{1/3}=\Gamma_{3002}$, $\Gamma^{1}_{2}=\Gamma_{3001}$, and $\Gamma^{2}_{2}=\Gamma_{3032}$. Hermitian matrices are $\Gamma_{\kappa \nu\rho\lambda}=\eta_\kappa \sigma_\nu \tau_\rho \beta_\lambda$, where $\{\eta_\kappa \}$, $\{\sigma_\kappa \}$, $\{\tau_\kappa \}$, $\{\beta_\kappa \}$ are Pauli matrices for $\kappa=0,\cdots, 3$, operating on the particle-hole, spin, valley and sublattice or layer indices, respectively. The Nambu spinor is $\Psi^{\top}_{\rm Nam}(\vec{k})=\left[ \Psi({\vec{k}}), \sigma_2 \tau_1 \beta_0 \Psi^\star(-{\vec{k}}) \right]$, where the eight-component spinor $\Psi^\top(\vec{k})=\left[\Psi_{\uparrow}(\vec{k}), \Psi_{\downarrow}(\vec{k}) \right]$ with $\sigma=\uparrow, \downarrow$ as two projections of electrons spin in the $z$ direction and $\Psi^{\top}_{\sigma}(\vec{k})=\left[ \Psi_{\sigma,+}(\vec{k}), \Psi_{\sigma,-}(\vec{k}) \right]$. Here $\top$ denotes transposition. For each spin projection, the two-component spinor near two opposite valleys at $\tau {\bf K}$ is defined as $\Psi_{\sigma,\tau}(\vec{k})=[A_{\sigma}(\tau {\bf K} + \vec{k}), B_{\sigma}(\tau {\bf K} + \vec{k})]$, where $\tau=\pm$. $A$ and $B$ are fermionic annihilation operator on the sites of two triangular sublattices of the honeycomb lattice. They, however, live on the top and bottom layers of BBLG and RTLG. Therefore, the sublattice and layer degrees of freedom are synonymous. Momentum $|\vec{k}| \ll |{\bf K}|$ is measured from the respective valley. We introduced the Nambu doubling to facilitate a forthcoming discussion on Majorana altermagnet. Until then, it is redundant. The Zeeman term ($\Delta_{\rm Z}$) is due to in-plane magnetic fields. In its absence, spherically symmetric energy spectra of $\hat{h}_{\ell}(\vec{k})$ are $\pm E_\ell(\vec{k})$, where $+$ ($-$) corresponds to the conduction (valence) band, and $E_{\ell}(\vec{k})=\alpha_\ell |\vec{k}|^\ell$~\cite{HJR2009, FanZhang2010, Vafek2010, BR2013, Szabo2021, Szabo2022}.

The effective Hamiltonian preserves the sublattice or layer ($S$) and valley ($T$) reflection symmetries, generated by $\Gamma_{0001}$ and $\Gamma_{0010}$, respectively, and accompanied by momentum reflections $\vec{k} \to (k_x,-k_y)$ and $\vec{k} \to (-k_x,k_y)$. Its time reversal symmetry is generated by ${\mathcal T}=\Gamma_{0210}{\mathcal K}$, where ${\mathcal K}$ is the complex conjugation and ${\mathcal T}^2=-1$. Thus electronic bands near two valleys are Kramer's (spin) degenerate. In the hole part of $\Psi_{\rm Nam}$, we absorb the unitary part of the time-reversal operator. The generator of spatial rotation is $\Gamma_{0033}$, and the low-energy Hamiltonian possesses a rotational symmetry, generated by $R_{\pi/2}=\exp[i \pi \Gamma_{0033}/4]$, when the momentum axes are rotated by an angle $\pi/(2 \ell)$. The U(1) translational symmetry is generated by $\Gamma_{0030}$. Light mass of carbon atoms allows us to neglect any spin-orbit coupling, and all the Hamiltonian are invariant under SU(2) spin rotation, generated by $\Gamma_{0s00}$ with $s=1,2,3$~\cite{HJR2009, Szabo2021, Szabo2022}. For details, see Sec.~S2 of the SM~\cite{SMaltermagnet}.

\emph{Spin nematicity}.~The underlying $D_{3d}$ group allows two spin-nematic orders transforming under the irreducible $E_g$ and $E_u$ representations. With respective amplitudes $\Delta_{E_g}$ and $\Delta_{E_u}$, effective single-particle Hamiltonian are  
\allowdisplaybreaks[4]
\begin{align}~\label{eq:spinnematic}
 \hat{h}^{\rm spin}_{E_g} \left(\Delta_{E_g}, \theta_{E_g} \right) &= \Delta_{E_g} \left[ \Gamma_{0301} \cos\theta_{E_g} - \Gamma_{0332} \sin \theta_{E_g} \right], \nonumber \\
\hat{h}^{\rm spin}_{E_u} \left(\Delta_{E_u}, \theta_{E_u} \right) &= \Delta_{E_u} \left[ \Gamma_{3331} \cos \theta_{E_u} - \Gamma_{3302} \sin \theta_{E_u} \right].
\end{align} 
The internal angles $\theta_{E_g}$ and $\theta_{E_u}$ are chosen spontaneously in the ordered states, detailed in Sec.~S3 of the SM~\cite{SMaltermagnet}, in which, without any loss of generality, the spin projection is picked in the $z$ direction. Negligibly small spin-orbit coupling allows classification of these orders solely in terms of the irreducible representation of the $D_{3d}$ group (ignoring the spin degrees of freedom),  without invoking spin space group. Two matrices of $\hat{h}^{\rm spin}_{E_g/E_u}$ constitute a vector under spatial rotation, generated by $\Gamma_{0033}$. So, the ordered states (with fixed $\theta_{E_g}$ or $\theta_{E_u}$) break the spatial rotational symmetry (yielding nematicity), while manifesting an invariance under $R_{\pi/2}$ rotation when $\cos\theta_{E_j} \to - \sin \theta_{E_j}$ and $\sin\theta_{E_j} \to  \cos \theta_{E_j}$ for $j=g,u$, thus satisfying the definition of altermagnets. The $E_g$ ($E_u$) spin nematicity breaks (preserves) the ${\mathcal T}$ symmetry, but preserves (breaks) the inversion or parity ${\mathcal P}$, generated by ${\mathcal P}=S T \equiv \Gamma_{0011}$ under which $\vec{k} \to -\vec{k}$. Thus only the $E_g$ spin-nematic order represents a conventional ${\mathcal T}$-odd altermagnet, while the $E_u$ counterpart corresponds to an odd-parity (${\mathcal P}$-odd) altermagnet. Nucleation of either order lifts the Kramer's degeneracy near each valley, discussed next. For lattice realizations of these orders see Fig.~\ref{fig:Fig2} and Sec.~S4 of the SM~\cite{SMaltermagnet}.

The reconstructed band structure with the onset of the spin nematic orders can be computed by diagonalizing
\begin{equation}~\label{eq:altmagHamil}
\hat{h}^{\rm alt}_j (\Delta_j, \theta_j) = \hat{h}_{\ell}(\vec{k}) + \hat{h}^{\rm spin}_j (\Delta_j, \theta_j),
\end{equation}
with $j=E_g$ and $E_u$. Near the $+{\bf K}$ valley, Kramers non-degenerate bands touch each other at Weyl points, located at $|\vec{k}|=\left([\Delta^2_{j}+\Delta^2_Z]^{1/2}/\alpha_{\ell} \right)^{1/\ell}$ and $\phi=(\theta_j+ m \pi)/\ell$. For spin-up ($\uparrow$) fermions, odd integer $m=1, \cdots, 2\ell-1$, while for spin-down ($\downarrow$) fermions even integer $m=0, \cdots, 2 \ell-2$. Therefore, for each spin projection, the linear band touching point of MLG shifts to a new position in the reciprocal space, whereas the bi-quadratic (bi-cubic) band touching point of BBLG (RTLG) splits into two (three) Weyl points around which the energy-momentum dispersion is linear. In the $E_g$ spin nematic phase, such a shift/splitting of the band touching points for spin-up (spin-down) fermions near $-{\bf K}$ valley is same as that of the spin-down (spin-up) fermions near $+{\bf K}$ valley in MLG and RTLG, but is identical for each spin projection near opposite valleys in BBLG. In the $E_u$ spin nematic state, this shift/splitting near the opposite valleys is identical in MLG and RTLG for each spin projection, whereas in BBLG such a shift/splitting near $-{\bf K}$ valley for spin-down (spin-up) fermions is same as that of the spin-up (spin-down) fermions near $+{\bf K}$ valley. The resulting reconstruction of electronic bands and its Kramers degeneracy lifting lead to altermagnetism in these spin nematic states, which we promote shortly. The distance between the Weyl nodes and the magnitude of the spin splitting of the Fermi surfaces with opposite spin projections are set by $\Delta_{E_g}$ and $\Delta_{E_u}$. 

The magnitudes of these two orders ($\Delta_{E_g}$ and $\Delta_{E_u}$) are expected to be a few meV, a typical scale of orderings in graphen heterostructures~\cite{Young2021I, Young2021II, Young2022, Pablo2022, Weitz2022}. However, their exact magnitude depends on the strength of corresponding four-fermion interaction, usually an unknown quantity in any interacting system, but expected to be a few eV in these systems~\cite{Katsnelson2011PRL}. Nevertheless, the local or momentum-independent nature of these orders make them energetically favored to be realized in real materials, supported by appropriate four-fermion interactions. In Sec.~S5 of the SM~\cite{SMaltermagnet}, we show their solutions as functions of the corresponding four-fermion interaction strength in the mean-field approximation. While in MLG these orderings take place beyond a critical strength of interaction due to vanishing density of states (DOS), in BBLG a constant DOS leads to their weak-coupling instabilities even for infinitesimal interactions. On the other hand, a diverging DOS causes strong nematic instabilities for weak interactions in RTLG.

\emph{Altermagnets}.~Emergent altermagnetism in the spin nematic phases can be recognized from the constant energy contours for opposite spin projections either in the valence or conduction band of the corresponding effective single-particle Hamiltonian [Eq.~\eqref{eq:spinnematic}]. The results are shown in Fig.~\ref{fig:Fig1} (left column). Such contours for spin-up and spin-down electrons do not overlap, but always enclose equal area in the reciprocal space (Fermi area). Thus these phases do not possess any net magnetic moment, despite lifting the Kramer's degeneracy from electronic bands. Hence, they represent altermagnets. Spin polarized constant energy contours cross each other at two, four and six points in MLG, BBLG, and RTLG, respectively. From the topology of such contours, it is evident that the same spin nematic order gives birth to $p$-wave, $d$-wave, and $f$-wave altermagnets in MLG, BBLG, and RTLG, respectively. Shortly, we will justify this claim quantitatively and attribute this emergent phenomena to the normal state band topology.

Application of a weak external in-plane magnetic field (no Landau quantization) splits the crossing points between contours belonging to opposite spin projections, where the $z$-component of electronic spin is zero, as shown in Fig.~\ref{fig:Fig1} (right column). The Zeeman coupling then takes place between the magnetic field and in-plane components (such as $x$) of electronic spin. The orbital effect of sufficiently weak in-plane magnetic fields is negligible in BBLG and RTLG in comparison to its Zeeman cousin, and is thus omitted here~\cite{RoyYang2013}. In-plane magnetic fields project the spin of altermagnets in the orthogonal easy-plane, and gap out the contour crossing points.

Classification of altermagnets, resulting from local spin nematic orders, in terms of the spherical harmonics is accomplished by casting their effective single-particle Hamiltonian [Eq.~\eqref{eq:spinnematic}] in the band electron basis. Then the kinetic energy term $\hat{h}^{\rm band}_\ell (\vec{k})=\alpha_\ell |\vec{k}|^\ell \bar{\Gamma}_{3003}-\Delta_{\rm Z} \bar{\Gamma}_{0100}$ becomes diagonal, achieved after a unitary rotation by $U$, constructed by columnwise arranging the eigenvectors of $\hat{h}_\ell(\vec{k})$ with $\Delta_{\rm Z}=0$. Here, $\bar{\Gamma}_{\kappa \nu\rho\lambda}=\eta_\kappa \sigma_\nu \tau_\rho \zeta_\lambda$ and the newly introduced Pauli matrices $\{ \zeta_\kappa \}$ operate on the band index (conduction and valence). In this basis, the local spin nematic orders from Eq.~\eqref{eq:spinnematic} take the form 
\begin{eqnarray}~\label{eq:bandprojection}
\hat{h}^{\rm spin}_{j,\rm band} &=& \Delta_{j} \bigg\{ \left[ \cos\theta_{j} \cos(\ell \phi) + \sin\theta_{j} \sin(\ell \phi) \right] \bar{\Gamma}^{j,\ell}_{\rm intra} \nonumber \\
&+& \left[ \cos\theta_{j} \sin(\ell \phi) - \sin\theta_{j} \cos(\ell \phi) \right] \bar{\Gamma}^{j,\ell}_{\rm inter} \bigg\}.
\end{eqnarray}    
The first (second) term captures the intraband (interband) component of the $j=E_g$ and $E_u$ nematic orders, ensured by the accompanying matrices taking the following form with $\kappa=3\; (0)$ for $j=E_g$ ($E_u$),
\allowdisplaybreaks[4]
\begin{eqnarray}
&&\bar{\Gamma}^{E_g,1/3}_{\rm intra}=\bar{\Gamma}^{E_u,2}_{\rm intra}=\bar{\Gamma}_{\kappa 333}, \:\:\:
\bar{\Gamma}^{E_g,2}_{\rm intra}=\bar{\Gamma}^{E_u,1/3}_{\rm intra}=\bar{\Gamma}_{\kappa 303} \nonumber \\
&&\bar{\Gamma}^{E_g,1/3}_{\rm inter}=\bar{\Gamma}^{E_u,2}_{\rm inter}=\bar{\Gamma}_{\kappa 302}, \:\:
\text{and} \:\:
\bar{\Gamma}^{E_g,2}_{\rm inter}=\bar{\Gamma}^{E_u,1/3}_{\rm inter}=\bar{\Gamma}_{\kappa 332}. \nonumber 
\end{eqnarray}
The intraband component is responsible for the topology of the constant energy contours (Fermi surface). From the definitions of cubic harmonics in two dimensions 
\begin{equation}
\left\{ \begin{array}{c}
\cos(\ell \phi) \\
\sin(\ell \phi)
\end{array}
\right\}
\propto Y^{-\ell}_{\ell} \left( \frac{\pi}{2}, \phi \right) 
\left\{ \begin{array}{c}
+ \\
-
\end{array}
\right\}
(-1)^\ell Y^{\ell}_{\ell} \left( \frac{\pi}{2}, \phi \right),  
\end{equation}
we identify that the altermagnets are $p$-wave, $d$-wave, and $f$-wave in nature in MLG ($\ell=1$), BBLG ($\ell=2$), and RTLG ($\ell=3$), respectively, resulting from their normal state band topology, described by the same harmonics.

\emph{Majorana altermagnet}.~As a penultimate topic, we showcase the possibility of realizing altermagnets of neutral Majorana fermions in local spin-triplet nematic superconductors. The $D_{3d}$ group allows only one such paired state, following the $E_u$ representation~\cite{BR2013, Szabo2021}, with the effective single-particle BdG Hamiltonian 
\begin{equation}
\hat{h}^{\rm pair}_{E_u} \left(\Delta^{p}_{E_u}, \theta^{p}_{E_u} \right) = \Delta^{p}_{E_u} \left[ \Gamma_{\alpha j31} \cos \theta^p_{E_u} - \Gamma_{\alpha j02} \sin \theta^p_{E_u} \right].
\end{equation} 
Here $\Delta^{p}_{E_u}$ is the pairing amplitude, $\theta^p_{E_u}$ determines the spatial orientation of Cooper pairs, and $\alpha=1,2$ reflects the U(1) gauge redundancy of the superconducting phase. For simplicity, we choose $\alpha=1$ and the Cooper pair spin in the $z$-direction ($j=3$). Microscopic origin of such pairing is shown Fig.~\ref{fig:Fig2} and in Sec.~S4 of the SM~\cite{SMaltermagnet}. The discussion for the $E_u$ spin nematic order directly applies here with the caveat that the Weyl nodes feature gapless Majorana excitations. We now enjoy the liberty to completely neglect the inter-band component of the pairing Hamiltonian [with $3 \to 1$ in the Nambu sector in Eq.~\eqref{eq:bandprojection}], as the effective attractive interaction exists only near the Fermi surface, found within the valence or conduction band upon doping these systems. Thus, the paired state also hosts altermagnets for neutral Majorana fermions [Fig.~\ref{fig:Fig1}]. We name them Majorana altermagnets. By the same analogy they are $p$-wave, $d$-wave, and $f$-wave in nature in MLG, BBLG, and RTLG, respectively.

\emph{Alter-valley magnet}.~Symmetry protected valley degree of freedom in graphene heterostructures enters their low-energy models as spin degrees of freedom [Eq.~\eqref{eq:freeHamil}], thus named pseudo-spin. We envision to construct altermagnetic states in terms of valley or pseudo-spin. Spin-up and spin-down components in this case translate into two valleys at $\pm {\bf K}$, and exchange of spin projections $\uparrow \leftrightarrow \downarrow$, leading to a change in spin angular momentum $S_z=\pm 2$ (in units of $\hbar/2$), maps onto ${\bf K} \leftrightarrow -{\bf K}$, causing a $2 {\bf K}$ momentum transfer. The proposed alter-valleymagnet is spin-disordered, and stems from the charge nematic orders, for which the effective single-particle Hamiltonian takes the form shown in Eq.~\eqref{eq:spinnematic}, with $3 \leftrightarrow 0$ in the Nambu sector and $3 \to 0$ in the spin sector of the corresponding $\Gamma$ matrices~\cite{Szabo2021}. A charge nematic phase then represents an alter-valley magnet if the displaced (in MLG) or distorted (in BBLG and RTLG) spin-degenerate Fermi surfaces near two inequivalent valleys do not map onto each other under a $2 {\bf K}$ translation (pseudo-spin flip). With this definition in hand, we recognize $E_g$ (in MLG and RTLG) and $E_u$ (in BBLG) charge nematic orders as alter-valleymagnet, with the corresponding spin degenerate Fermi surfaces from the opposite valleys shown in a single frame in Fig.~\ref{fig:Fig1} (left column), where $\uparrow/\downarrow \; \equiv +/-{\bf K}$. Lattice realizations of these orders are shown in Fig.~\ref{fig:Fig2} and Sec.~S4 of the SM~\cite{SMaltermagnet}. They can be identified from anisotropic charge transport (signature of nematicity), combined with ARPES measurement, confirming unnested Fermi surfaces under $2{\bf K}$ momentum exchange~\cite{ARPESGraphene}.

\emph{Summary and discussions}.~We show that the band topology of non-interacting electrons plays a decisive role in determining the geometry of emergent altermagnets from the local spin-nematic orders. As examples, we consider graphene-based crystalline heterostructures, namely MLG, BBLG, and RTLG, displaying linear, quadratic, and cubic band dispersion, captured by $\ell=1,2$, and $3$ harmonics, respectively. As a result, the local spin-nematic orders foster $p$-wave, $d$-wave, and $f$-wave altermagnets, respectively, inheriting their geometry from the free fermion band topology. The same conclusions hold in a spin-triplet nematic local superconductor, harboring Majorana altermagnets. In addition, the valley or pseudo-spin degree of freedom allows us to unfold the possibility of spin-disordered alter-valley magnetic phases. Present discussion opens up various fascinating future directions, among which generalizing these concepts to strong spin-orbit coupled and three-dimensional materials, emergent superconductors in doped altermagnets~\cite{Brekke2023, Li2023, Zhu2023, Zhang2024, Ghorashi2023, Chakraborty2023} are the prominent ones. The spin nematicity driven emergent altermagnetism, yielding reconstructed spin non-degenerate band structure (Fig.~\ref{fig:Fig1}) can be identified from spin-resolved ARPES and fast Fourier transformed STM measurements. The broken ${\mathcal T}$ symmetry in altermagnets should also be verified from Faraday and Kerr rotations~\cite{Kapitulnik2009}.

Topological quantum chemistry nowadays is routinely employed to mine quantum materials with unusual electronic band dispersion~\cite{Slager2017, bernevig2017, Vishwanath2017NatComm, Zhang2019, Vergniory2019, Tang2019}. Our proposed symmetry-based sufficiently general one-to-one correspondence between band topology and altermagnet geometry should therefore open an unexplored and fascinating avenue to harness these exotic quantum magnets in a predictive way. Within the landscape of graphene heterostructures recent experiments have unveiled several ordered phases (including superconductors) in doped BBLG and RTLG, when the layer-inversion symmetry is broken by an external displacement electric field~\cite{Young2021I, Young2021II, Young2022, Pablo2022, Weitz2022}. In biased BBLG and RTLG, spin-triplet superconductivity has been observed~\cite{Young2021II, Young2022}. Their nematic nature can now be established from direction and spin  dependent longitudinal thermal transport measurement, unfolding the proposed Majorana altermagnet. By contrast, spin nematic orders, yielding altermagnets, can be pinned from direction and spin dependent longitudinal regular charge transport measurements. The magnitude of the spin-resolved longitudinal conductivity (electrical or thermal) along a specific direction is proportional to the Fermi vector along it (see Fig.~\ref{fig:Fig1}). While nematic orders have also been observed experimentally in BBLG, their spin orientation has remained unexplored so far~\cite{novoselov:nematic, yacoby:nematic}. As new phases in the global phase diagram of graphene-based crystalline (non-Moir\'e) systems are still being discovered, they constitute a promising material platform, where our predicted altermagnets, including their Majorana and valley cousins, can in principle be observed, stimulating new experiments in this direction. Our predictions are not affected by the trigonal warping in BBLG and RTLG, as shown in Sec.~S6 of the SM~\cite{SMaltermagnet}.

\emph{Acknowledgments}.~S.K.D.\ was supported by the Startup Grant of B.R.\ from Lehigh University. B.R.\ was supported by NSF CAREER Grant No.\ DMR-2238679. We thank Daniel Salib for comments on the manuscript.

\bibliography{ref_altermagnet}

\begin{thebibliography}{75}%
\makeatletter
\providecommand \@ifxundefined [1]{%
 \@ifx{#1\undefined}
}%
\providecommand \@ifnum [1]{%
 \ifnum #1\expandafter \@firstoftwo
 \else \expandafter \@secondoftwo
 \fi
}%
\providecommand \@ifx [1]{%
 \ifx #1\expandafter \@firstoftwo
 \else \expandafter \@secondoftwo
 \fi
}%
\providecommand \natexlab [1]{#1}%
\providecommand \enquote  [1]{``#1''}%
\providecommand \bibnamefont  [1]{#1}%
\providecommand \bibfnamefont [1]{#1}%
\providecommand \citenamefont [1]{#1}%
\providecommand \href@noop [0]{\@secondoftwo}%
\providecommand \href [0]{\begingroup \@sanitize@url \@href}%
\providecommand \@href[1]{\@@startlink{#1}\@@href}%
\providecommand \@@href[1]{\endgroup#1\@@endlink}%
\providecommand \@sanitize@url [0]{\catcode `\\12\catcode `\$12\catcode
  `\&12\catcode `\#12\catcode `\^12\catcode `\_12\catcode `\%12\relax}%
\providecommand \@@startlink[1]{}%
\providecommand \@@endlink[0]{}%
\providecommand \url  [0]{\begingroup\@sanitize@url \@url }%
\providecommand \@url [1]{\endgroup\@href {#1}{\urlprefix }}%
\providecommand \urlprefix  [0]{URL }%
\providecommand \Eprint [0]{\href }%
\providecommand \doibase [0]{https://doi.org/}%
\providecommand \selectlanguage [0]{\@gobble}%
\providecommand \bibinfo  [0]{\@secondoftwo}%
\providecommand \bibfield  [0]{\@secondoftwo}%
\providecommand \translation [1]{[#1]}%
\providecommand \BibitemOpen [0]{}%
\providecommand \bibitemStop [0]{}%
\providecommand \bibitemNoStop [0]{.\EOS\space}%
\providecommand \EOS [0]{\spacefactor3000\relax}%
\providecommand \BibitemShut  [1]{\csname bibitem#1\endcsname}%
\let\auto@bib@innerbib\@empty
\bibitem [{\citenamefont {Hayami}\ \emph {et~al.}(2019)\citenamefont {Hayami},
  \citenamefont {Yanagi},\ and\ \citenamefont {Kusunose}}]{Hayami2019}%
  \BibitemOpen
  \bibfield  {author} {\bibinfo {author} {\bibfnamefont {S.}~\bibnamefont
  {Hayami}}, \bibinfo {author} {\bibfnamefont {Y.}~\bibnamefont {Yanagi}},\
  and\ \bibinfo {author} {\bibfnamefont {H.}~\bibnamefont {Kusunose}},\
  }\bibfield  {title} {\bibinfo {title} {{Momentum-Dependent Spin Splitting by
  Collinear Antiferromagnetic Ordering}},\ }\href
  {https://doi.org/10.7566/jpsj.88.123702} {\bibfield  {journal} {\bibinfo
  {journal} {J. Phys. Soc. Jpn.}\ }\textbf {\bibinfo {volume} {88}},\ \bibinfo
  {pages} {123702} (\bibinfo {year} {2019})}\BibitemShut {NoStop}%
\bibitem [{\citenamefont {Ahn}\ \emph {et~al.}(2019)\citenamefont {Ahn},
  \citenamefont {Hariki}, \citenamefont {Lee},\ and\ \citenamefont
  {Kune\ifmmode~\check{s}\else \v{s}\fi{}}}]{Ahn2019}%
  \BibitemOpen
  \bibfield  {author} {\bibinfo {author} {\bibfnamefont {K.-H.}\ \bibnamefont
  {Ahn}}, \bibinfo {author} {\bibfnamefont {A.}~\bibnamefont {Hariki}},
  \bibinfo {author} {\bibfnamefont {K.-W.}\ \bibnamefont {Lee}},\ and\ \bibinfo
  {author} {\bibfnamefont {J.}~\bibnamefont {Kune\ifmmode~\check{s}\else
  \v{s}\fi{}}},\ }\bibfield  {title} {\bibinfo {title} {{Antiferromagnetism in
  ${\mathrm{RuO}}_{2}$ as $d$-wave Pomeranchuk instability}},\ }\href
  {https://doi.org/10.1103/PhysRevB.99.184432} {\bibfield  {journal} {\bibinfo
  {journal} {Phys. Rev. B}\ }\textbf {\bibinfo {volume} {99}},\ \bibinfo
  {pages} {184432} (\bibinfo {year} {2019})}\BibitemShut {NoStop}%
\bibitem [{\citenamefont {{\v{S}}mejkal}\ \emph {et~al.}(2020)\citenamefont
  {{\v{S}}mejkal}, \citenamefont {Gonz{\'a}lez-Hern{\'a}ndez}, \citenamefont
  {Jungwirth},\ and\ \citenamefont {Sinova}}]{Smejkal2020}%
  \BibitemOpen
  \bibfield  {author} {\bibinfo {author} {\bibfnamefont {L.}~\bibnamefont
  {{\v{S}}mejkal}}, \bibinfo {author} {\bibfnamefont {R.}~\bibnamefont
  {Gonz{\'a}lez-Hern{\'a}ndez}}, \bibinfo {author} {\bibfnamefont
  {T.}~\bibnamefont {Jungwirth}},\ and\ \bibinfo {author} {\bibfnamefont
  {J.}~\bibnamefont {Sinova}},\ }\bibfield  {title} {\bibinfo {title} {{Crystal
  time-reversal symmetry breaking and spontaneous {Hall} effect in collinear
  antiferromagnets}},\ }\href {https://doi.org/10.1126/sciadv.aaz8809}
  {\bibfield  {journal} {\bibinfo  {journal} {Sci. Adv.}\ }\textbf {\bibinfo
  {volume} {6}},\ \bibinfo {pages} {eaaz8809} (\bibinfo {year}
  {2020})}\BibitemShut {NoStop}%
\bibitem [{\citenamefont {Chen}\ \emph {et~al.}(2020)\citenamefont {Chen},
  \citenamefont {Wang}, \citenamefont {Xiao}, \citenamefont {Guo},
  \citenamefont {Niu},\ and\ \citenamefont {MacDonald}}]{Chen2020}%
  \BibitemOpen
  \bibfield  {author} {\bibinfo {author} {\bibfnamefont {H.}~\bibnamefont
  {Chen}}, \bibinfo {author} {\bibfnamefont {T.-C.}\ \bibnamefont {Wang}},
  \bibinfo {author} {\bibfnamefont {D.}~\bibnamefont {Xiao}}, \bibinfo {author}
  {\bibfnamefont {G.-Y.}\ \bibnamefont {Guo}}, \bibinfo {author} {\bibfnamefont
  {Q.}~\bibnamefont {Niu}},\ and\ \bibinfo {author} {\bibfnamefont {A.~H.}\
  \bibnamefont {MacDonald}},\ }\bibfield  {title} {\bibinfo {title}
  {{Manipulating anomalous Hall antiferromagnets with magnetic fields}},\
  }\href {https://doi.org/10.1103/PhysRevB.101.104418} {\bibfield  {journal}
  {\bibinfo  {journal} {Phys. Rev. B}\ }\textbf {\bibinfo {volume} {101}},\
  \bibinfo {pages} {104418} (\bibinfo {year} {2020})}\BibitemShut {NoStop}%
\bibitem [{\citenamefont {Ma}\ \emph {et~al.}(2021)\citenamefont {Ma},
  \citenamefont {Hu}, \citenamefont {Li}, \citenamefont {Liu}, \citenamefont
  {Yao}, \citenamefont {Jia},\ and\ \citenamefont {Liu}}]{JunweiLiu2021}%
  \BibitemOpen
  \bibfield  {author} {\bibinfo {author} {\bibfnamefont {H.-Y.}\ \bibnamefont
  {Ma}}, \bibinfo {author} {\bibfnamefont {M.}~\bibnamefont {Hu}}, \bibinfo
  {author} {\bibfnamefont {N.}~\bibnamefont {Li}}, \bibinfo {author}
  {\bibfnamefont {J.}~\bibnamefont {Liu}}, \bibinfo {author} {\bibfnamefont
  {W.}~\bibnamefont {Yao}}, \bibinfo {author} {\bibfnamefont {J.-F.}\
  \bibnamefont {Jia}},\ and\ \bibinfo {author} {\bibfnamefont {J.}~\bibnamefont
  {Liu}},\ }\bibfield  {title} {\bibinfo {title} {{Multifunctional
  antiferromagnetic materials with giant piezomagnetism and noncollinear spin
  current}},\ }\href {https://doi.org/10.1038/s41467-021-23127-7} {\bibfield
  {journal} {\bibinfo  {journal} {Nat. Commun.}\ }\textbf {\bibinfo {volume}
  {12}},\ \bibinfo {pages} {2846} (\bibinfo {year} {2021})}\BibitemShut
  {NoStop}%
\bibitem [{\citenamefont {\ifmmode~\check{S}\else \v{S}\fi{}mejkal}\ \emph
  {et~al.}(2022{\natexlab{a}})\citenamefont {\ifmmode~\check{S}\else
  \v{S}\fi{}mejkal}, \citenamefont {Sinova},\ and\ \citenamefont
  {Jungwirth}}]{Smejkal2022I}%
  \BibitemOpen
  \bibfield  {author} {\bibinfo {author} {\bibfnamefont {L.}~\bibnamefont
  {\ifmmode~\check{S}\else \v{S}\fi{}mejkal}}, \bibinfo {author} {\bibfnamefont
  {J.}~\bibnamefont {Sinova}},\ and\ \bibinfo {author} {\bibfnamefont
  {T.}~\bibnamefont {Jungwirth}},\ }\bibfield  {title} {\bibinfo {title}
  {{Emerging Research Landscape of Altermagnetism}},\ }\href
  {https://doi.org/10.1103/PhysRevX.12.040501} {\bibfield  {journal} {\bibinfo
  {journal} {Phys. Rev. X}\ }\textbf {\bibinfo {volume} {12}},\ \bibinfo
  {pages} {040501} (\bibinfo {year} {2022}{\natexlab{a}})}\BibitemShut
  {NoStop}%
\bibitem [{\citenamefont {\ifmmode~\check{S}\else \v{S}\fi{}mejkal}\ \emph
  {et~al.}(2022{\natexlab{b}})\citenamefont {\ifmmode~\check{S}\else
  \v{S}\fi{}mejkal}, \citenamefont {Sinova},\ and\ \citenamefont
  {Jungwirth}}]{Smejkal2022II}%
  \BibitemOpen
  \bibfield  {author} {\bibinfo {author} {\bibfnamefont {L.}~\bibnamefont
  {\ifmmode~\check{S}\else \v{S}\fi{}mejkal}}, \bibinfo {author} {\bibfnamefont
  {J.}~\bibnamefont {Sinova}},\ and\ \bibinfo {author} {\bibfnamefont
  {T.}~\bibnamefont {Jungwirth}},\ }\bibfield  {title} {\bibinfo {title}
  {{Beyond Conventional Ferromagnetism and Antiferromagnetism: A Phase with
  Nonrelativistic Spin and Crystal Rotation Symmetry}},\ }\href
  {https://doi.org/10.1103/PhysRevX.12.031042} {\bibfield  {journal} {\bibinfo
  {journal} {Phys. Rev. X}\ }\textbf {\bibinfo {volume} {12}},\ \bibinfo
  {pages} {031042} (\bibinfo {year} {2022}{\natexlab{b}})}\BibitemShut
  {NoStop}%
\bibitem [{\citenamefont {{\v{S}}mejkal}\ \emph {et~al.}(2022)\citenamefont
  {{\v{S}}mejkal}, \citenamefont {{MacDonald}}, \citenamefont {Sinova},
  \citenamefont {Nakatsuji},\ and\ \citenamefont {Jungwirth}}]{Smejkal2022III}%
  \BibitemOpen
  \bibfield  {author} {\bibinfo {author} {\bibfnamefont {L.}~\bibnamefont
  {{\v{S}}mejkal}}, \bibinfo {author} {\bibfnamefont {A.~H.}\ \bibnamefont
  {{MacDonald}}}, \bibinfo {author} {\bibfnamefont {J.}~\bibnamefont {Sinova}},
  \bibinfo {author} {\bibfnamefont {S.}~\bibnamefont {Nakatsuji}},\ and\
  \bibinfo {author} {\bibfnamefont {T.}~\bibnamefont {Jungwirth}},\ }\bibfield
  {title} {\bibinfo {title} {{Anomalous Hall antiferromagnets}},\ }\href
  {https://doi.org/10.1038/s41578-022-00430-3} {\bibfield  {journal} {\bibinfo
  {journal} {Nat. Rev. Mater.}\ }\textbf {\bibinfo {volume} {7}},\ \bibinfo
  {pages} {482} (\bibinfo {year} {2022})}\BibitemShut {NoStop}%
\bibitem [{\citenamefont {Turek}(2022)}]{Turek2022}%
  \BibitemOpen
  \bibfield  {author} {\bibinfo {author} {\bibfnamefont {I.}~\bibnamefont
  {Turek}},\ }\bibfield  {title} {\bibinfo {title} {{Altermagnetism and
  magnetic groups with pseudoscalar electron spin}},\ }\href
  {https://doi.org/10.1103/PhysRevB.106.094432} {\bibfield  {journal} {\bibinfo
   {journal} {Phys. Rev. B}\ }\textbf {\bibinfo {volume} {106}},\ \bibinfo
  {pages} {094432} (\bibinfo {year} {2022})}\BibitemShut {NoStop}%
\bibitem [{\citenamefont {\ifmmode~\check{S}\else \v{S}\fi{}mejkal}\ \emph
  {et~al.}(2023)\citenamefont {\ifmmode~\check{S}\else \v{S}\fi{}mejkal},
  \citenamefont {Marmodoro}, \citenamefont {Ahn}, \citenamefont
  {Gonz\'alez-Hern\'andez}, \citenamefont {Turek}, \citenamefont {Mankovsky},
  \citenamefont {Ebert}, \citenamefont {D'Souza}, \citenamefont
  {\ifmmode~\check{S}\else \v{S}\fi{}ipr}, \citenamefont {Sinova},\ and\
  \citenamefont {Jungwirth}}]{Libor2023}%
  \BibitemOpen
  \bibfield  {author} {\bibinfo {author} {\bibfnamefont {L.}~\bibnamefont
  {\ifmmode~\check{S}\else \v{S}\fi{}mejkal}}, \bibinfo {author} {\bibfnamefont
  {A.}~\bibnamefont {Marmodoro}}, \bibinfo {author} {\bibfnamefont {K.-H.}\
  \bibnamefont {Ahn}}, \bibinfo {author} {\bibfnamefont {R.}~\bibnamefont
  {Gonz\'alez-Hern\'andez}}, \bibinfo {author} {\bibfnamefont {I.}~\bibnamefont
  {Turek}}, \bibinfo {author} {\bibfnamefont {S.}~\bibnamefont {Mankovsky}},
  \bibinfo {author} {\bibfnamefont {H.}~\bibnamefont {Ebert}}, \bibinfo
  {author} {\bibfnamefont {S.~W.}\ \bibnamefont {D'Souza}}, \bibinfo {author}
  {\bibfnamefont {O.~c.~v.}\ \bibnamefont {\ifmmode~\check{S}\else
  \v{S}\fi{}ipr}}, \bibinfo {author} {\bibfnamefont {J.}~\bibnamefont
  {Sinova}},\ and\ \bibinfo {author} {\bibfnamefont {T.~c.~v.}\ \bibnamefont
  {Jungwirth}},\ }\bibfield  {title} {\bibinfo {title} {{Chiral Magnons in
  Altermagnetic ${\mathrm{RuO}}_{2}$}},\ }\href
  {https://doi.org/10.1103/PhysRevLett.131.256703} {\bibfield  {journal}
  {\bibinfo  {journal} {Phys. Rev. Lett.}\ }\textbf {\bibinfo {volume} {131}},\
  \bibinfo {pages} {256703} (\bibinfo {year} {2023})}\BibitemShut {NoStop}%
\bibitem [{\citenamefont {Mazin}(2023)}]{Mazin2023}%
  \BibitemOpen
  \bibfield  {author} {\bibinfo {author} {\bibfnamefont {I.~I.}\ \bibnamefont
  {Mazin}},\ }\bibfield  {title} {\bibinfo {title} {{Altermagnetism in MnTe:
  Origin, predicted manifestations, and routes to detwinning}},\ }\href
  {https://doi.org/10.1103/PhysRevB.107.L100418} {\bibfield  {journal}
  {\bibinfo  {journal} {Phys. Rev. B}\ }\textbf {\bibinfo {volume} {107}},\
  \bibinfo {pages} {L100418} (\bibinfo {year} {2023})}\BibitemShut {NoStop}%
\bibitem [{\citenamefont {Fakhredine}\ \emph {et~al.}(2023)\citenamefont
  {Fakhredine}, \citenamefont {Sattigeri}, \citenamefont {Cuono},\ and\
  \citenamefont {Autieri}}]{Fakhredine2023}%
  \BibitemOpen
  \bibfield  {author} {\bibinfo {author} {\bibfnamefont {A.}~\bibnamefont
  {Fakhredine}}, \bibinfo {author} {\bibfnamefont {R.~M.}\ \bibnamefont
  {Sattigeri}}, \bibinfo {author} {\bibfnamefont {G.}~\bibnamefont {Cuono}},\
  and\ \bibinfo {author} {\bibfnamefont {C.}~\bibnamefont {Autieri}},\
  }\bibfield  {title} {\bibinfo {title} {{Interplay between altermagnetism and
  nonsymmorphic symmetries generating large anomalous Hall conductivity by
  semi-Dirac points induced anticrossings}},\ }\href
  {https://doi.org/10.1103/PhysRevB.108.115138} {\bibfield  {journal} {\bibinfo
   {journal} {Phys. Rev. B}\ }\textbf {\bibinfo {volume} {108}},\ \bibinfo
  {pages} {115138} (\bibinfo {year} {2023})}\BibitemShut {NoStop}%
\bibitem [{\citenamefont {Gao}\ \emph {et~al.}(2025)\citenamefont {Gao},
  \citenamefont {Qu}, \citenamefont {Zeng}, \citenamefont {Liu}, \citenamefont
  {Wen}, \citenamefont {Sun}, \citenamefont {Guo},\ and\ \citenamefont
  {Lu}}]{Gao2023}%
  \BibitemOpen
  \bibfield  {author} {\bibinfo {author} {\bibfnamefont {Z.-F.}\ \bibnamefont
  {Gao}}, \bibinfo {author} {\bibfnamefont {S.}~\bibnamefont {Qu}}, \bibinfo
  {author} {\bibfnamefont {B.}~\bibnamefont {Zeng}}, \bibinfo {author}
  {\bibfnamefont {Y.}~\bibnamefont {Liu}}, \bibinfo {author} {\bibfnamefont
  {J.-R.}\ \bibnamefont {Wen}}, \bibinfo {author} {\bibfnamefont
  {H.}~\bibnamefont {Sun}}, \bibinfo {author} {\bibfnamefont {P.-J.}\
  \bibnamefont {Guo}},\ and\ \bibinfo {author} {\bibfnamefont {Z.-Y.}\
  \bibnamefont {Lu}},\ }\bibfield  {title} {\bibinfo {title} {{AI-accelerated
  discovery of altermagnetic materials}},\ }\href
  {https://doi.org/10.1093/nsr/nwaf066} {\bibfield  {journal} {\bibinfo
  {journal} {Natl. Sci. Rev.}\ }\textbf {\bibinfo {volume} {12}},\ \bibinfo
  {pages} {nwaf066} (\bibinfo {year} {2025})}\BibitemShut {NoStop}%
\bibitem [{\citenamefont {Zhang}\ \emph
  {et~al.}(2024{\natexlab{a}})\citenamefont {Zhang}, \citenamefont {Cui},
  \citenamefont {Li}, \citenamefont {Duan}, \citenamefont {Li}, \citenamefont
  {Yu},\ and\ \citenamefont {Yao}}]{Run2023}%
  \BibitemOpen
  \bibfield  {author} {\bibinfo {author} {\bibfnamefont {R.-W.}\ \bibnamefont
  {Zhang}}, \bibinfo {author} {\bibfnamefont {C.}~\bibnamefont {Cui}}, \bibinfo
  {author} {\bibfnamefont {R.}~\bibnamefont {Li}}, \bibinfo {author}
  {\bibfnamefont {J.}~\bibnamefont {Duan}}, \bibinfo {author} {\bibfnamefont
  {L.}~\bibnamefont {Li}}, \bibinfo {author} {\bibfnamefont {Z.-M.}\
  \bibnamefont {Yu}},\ and\ \bibinfo {author} {\bibfnamefont {Y.}~\bibnamefont
  {Yao}},\ }\bibfield  {title} {\bibinfo {title} {{Predictable Gate-Field
  Control of Spin in Altermagnets with Spin-Layer Coupling}},\ }\href
  {https://doi.org/10.1103/PhysRevLett.133.056401} {\bibfield  {journal}
  {\bibinfo  {journal} {Phys. Rev. Lett.}\ }\textbf {\bibinfo {volume} {133}},\
  \bibinfo {pages} {056401} (\bibinfo {year} {2024}{\natexlab{a}})}\BibitemShut
  {NoStop}%
\bibitem [{\citenamefont {Chi}\ \emph {et~al.}(2024)\citenamefont {Chi},
  \citenamefont {Jiang}, \citenamefont {Zhu}, \citenamefont {Yu}, \citenamefont
  {Wan}, \citenamefont {Zhang},\ and\ \citenamefont {Han}}]{Chi2023}%
  \BibitemOpen
  \bibfield  {author} {\bibinfo {author} {\bibfnamefont {B.}~\bibnamefont
  {Chi}}, \bibinfo {author} {\bibfnamefont {L.}~\bibnamefont {Jiang}}, \bibinfo
  {author} {\bibfnamefont {Y.}~\bibnamefont {Zhu}}, \bibinfo {author}
  {\bibfnamefont {G.}~\bibnamefont {Yu}}, \bibinfo {author} {\bibfnamefont
  {C.}~\bibnamefont {Wan}}, \bibinfo {author} {\bibfnamefont {J.}~\bibnamefont
  {Zhang}},\ and\ \bibinfo {author} {\bibfnamefont {X.}~\bibnamefont {Han}},\
  }\bibfield  {title} {\bibinfo {title} {{Crystal-facet-oriented altermagnets
  for detecting ferromagnetic and antiferromagnetic states by giant tunneling
  magnetoresistance}},\ }\href
  {https://doi.org/10.1103/PhysRevApplied.21.034038} {\bibfield  {journal}
  {\bibinfo  {journal} {Phys. Rev. Appl.}\ }\textbf {\bibinfo {volume} {21}},\
  \bibinfo {pages} {034038} (\bibinfo {year} {2024})}\BibitemShut {NoStop}%
\bibitem [{\citenamefont {Ouassou}\ \emph {et~al.}(2023)\citenamefont
  {Ouassou}, \citenamefont {Brataas},\ and\ \citenamefont
  {Linder}}]{Ouassou2023}%
  \BibitemOpen
  \bibfield  {author} {\bibinfo {author} {\bibfnamefont {J.~A.}\ \bibnamefont
  {Ouassou}}, \bibinfo {author} {\bibfnamefont {A.}~\bibnamefont {Brataas}},\
  and\ \bibinfo {author} {\bibfnamefont {J.}~\bibnamefont {Linder}},\
  }\bibfield  {title} {\bibinfo {title} {dc josephson effect in altermagnets},\
  }\href {https://doi.org/10.1103/PhysRevLett.131.076003} {\bibfield  {journal}
  {\bibinfo  {journal} {Phys. Rev. Lett.}\ }\textbf {\bibinfo {volume} {131}},\
  \bibinfo {pages} {076003} (\bibinfo {year} {2023})}\BibitemShut {NoStop}%
\bibitem [{\citenamefont {Steward}\ \emph {et~al.}(2023)\citenamefont
  {Steward}, \citenamefont {Fernandes},\ and\ \citenamefont
  {Schmalian}}]{Steward2023}%
  \BibitemOpen
  \bibfield  {author} {\bibinfo {author} {\bibfnamefont {C.~R.~W.}\
  \bibnamefont {Steward}}, \bibinfo {author} {\bibfnamefont {R.~M.}\
  \bibnamefont {Fernandes}},\ and\ \bibinfo {author} {\bibfnamefont
  {J.}~\bibnamefont {Schmalian}},\ }\bibfield  {title} {\bibinfo {title}
  {Dynamic paramagnon-polarons in altermagnets},\ }\href
  {https://doi.org/10.1103/PhysRevB.108.144418} {\bibfield  {journal} {\bibinfo
   {journal} {Phys. Rev. B}\ }\textbf {\bibinfo {volume} {108}},\ \bibinfo
  {pages} {144418} (\bibinfo {year} {2023})}\BibitemShut {NoStop}%
\bibitem [{\citenamefont {Fernandes}\ \emph {et~al.}(2024)\citenamefont
  {Fernandes}, \citenamefont {de~Carvalho}, \citenamefont {Birol},\ and\
  \citenamefont {Pereira}}]{Fernandes2024}%
  \BibitemOpen
  \bibfield  {author} {\bibinfo {author} {\bibfnamefont {R.~M.}\ \bibnamefont
  {Fernandes}}, \bibinfo {author} {\bibfnamefont {V.~S.}\ \bibnamefont
  {de~Carvalho}}, \bibinfo {author} {\bibfnamefont {T.}~\bibnamefont {Birol}},\
  and\ \bibinfo {author} {\bibfnamefont {R.~G.}\ \bibnamefont {Pereira}},\
  }\bibfield  {title} {\bibinfo {title} {Topological transition from nodal to
  nodeless zeeman splitting in altermagnets},\ }\href
  {https://doi.org/10.1103/PhysRevB.109.024404} {\bibfield  {journal} {\bibinfo
   {journal} {Phys. Rev. B}\ }\textbf {\bibinfo {volume} {109}},\ \bibinfo
  {pages} {024404} (\bibinfo {year} {2024})}\BibitemShut {NoStop}%
\bibitem [{\citenamefont {Roig}\ \emph {et~al.}(2024)\citenamefont {Roig},
  \citenamefont {Kreisel}, \citenamefont {Yu}, \citenamefont {Andersen},\ and\
  \citenamefont {Agterberg}}]{Agterberg2024}%
  \BibitemOpen
  \bibfield  {author} {\bibinfo {author} {\bibfnamefont {M.}~\bibnamefont
  {Roig}}, \bibinfo {author} {\bibfnamefont {A.}~\bibnamefont {Kreisel}},
  \bibinfo {author} {\bibfnamefont {Y.}~\bibnamefont {Yu}}, \bibinfo {author}
  {\bibfnamefont {B.~M.}\ \bibnamefont {Andersen}},\ and\ \bibinfo {author}
  {\bibfnamefont {D.~F.}\ \bibnamefont {Agterberg}},\ }\bibfield  {title}
  {\bibinfo {title} {{Minimal models for altermagnetism}},\ }\href
  {https://doi.org/10.1103/PhysRevB.110.144412} {\bibfield  {journal} {\bibinfo
   {journal} {Phys. Rev. B}\ }\textbf {\bibinfo {volume} {110}},\ \bibinfo
  {pages} {144412} (\bibinfo {year} {2024})}\BibitemShut {NoStop}%
\bibitem [{\citenamefont {Parshukov}\ \emph {et~al.}()\citenamefont
  {Parshukov}, \citenamefont {Wiedmann},\ and\ \citenamefont
  {Schnyder}}]{Parshukov2024}%
  \BibitemOpen
  \bibfield  {author} {\bibinfo {author} {\bibfnamefont {K.}~\bibnamefont
  {Parshukov}}, \bibinfo {author} {\bibfnamefont {R.}~\bibnamefont
  {Wiedmann}},\ and\ \bibinfo {author} {\bibfnamefont {A.~P.}\ \bibnamefont
  {Schnyder}},\ }\href@noop {} {\bibinfo {title} {{Topological responses from
  gapped Weyl points in 2D altermagnets}}},\ \Eprint
  {https://arxiv.org/abs/arXiv:2403.09520} {arXiv:2403.09520} \BibitemShut
  {NoStop}%
\bibitem [{\citenamefont {Feng}\ \emph {et~al.}(2022)\citenamefont {Feng},
  \citenamefont {Zhou}, \citenamefont {Smejkal}, \citenamefont {Wu},
  \citenamefont {Zhu}, \citenamefont {Guo}, \citenamefont {Gonzalez-Hernandez},
  \citenamefont {Wang}, \citenamefont {Yan}, \citenamefont {Qin}, \citenamefont
  {Zhang}, \citenamefont {Wu}, \citenamefont {Chen}, \citenamefont {Meng},
  \citenamefont {Liu}, \citenamefont {Xia}, \citenamefont {Sinova},
  \citenamefont {Jungwirth},\ and\ \citenamefont {Liu}}]{Feng2022}%
  \BibitemOpen
  \bibfield  {author} {\bibinfo {author} {\bibfnamefont {Z.}~\bibnamefont
  {Feng}}, \bibinfo {author} {\bibfnamefont {X.}~\bibnamefont {Zhou}}, \bibinfo
  {author} {\bibfnamefont {L.}~\bibnamefont {Smejkal}}, \bibinfo {author}
  {\bibfnamefont {L.}~\bibnamefont {Wu}}, \bibinfo {author} {\bibfnamefont
  {Z.}~\bibnamefont {Zhu}}, \bibinfo {author} {\bibfnamefont {H.}~\bibnamefont
  {Guo}}, \bibinfo {author} {\bibfnamefont {R.}~\bibnamefont
  {Gonzalez-Hernandez}}, \bibinfo {author} {\bibfnamefont {X.}~\bibnamefont
  {Wang}}, \bibinfo {author} {\bibfnamefont {H.}~\bibnamefont {Yan}}, \bibinfo
  {author} {\bibfnamefont {P.}~\bibnamefont {Qin}}, \bibinfo {author}
  {\bibfnamefont {X.}~\bibnamefont {Zhang}}, \bibinfo {author} {\bibfnamefont
  {H.}~\bibnamefont {Wu}}, \bibinfo {author} {\bibfnamefont {H.}~\bibnamefont
  {Chen}}, \bibinfo {author} {\bibfnamefont {Z.}~\bibnamefont {Meng}}, \bibinfo
  {author} {\bibfnamefont {L.}~\bibnamefont {Liu}}, \bibinfo {author}
  {\bibfnamefont {Z.}~\bibnamefont {Xia}}, \bibinfo {author} {\bibfnamefont
  {J.}~\bibnamefont {Sinova}}, \bibinfo {author} {\bibfnamefont
  {T.}~\bibnamefont {Jungwirth}},\ and\ \bibinfo {author} {\bibfnamefont
  {Z.}~\bibnamefont {Liu}},\ }\bibfield  {title} {\bibinfo {title} {{An
  anomalous Hall effect in altermagnetic ruthenium dioxide}},\ }\href
  {https://doi.org/10.1038/s41928-022-00866-z} {\bibfield  {journal} {\bibinfo
  {journal} {Nat. Electron.}\ }\textbf {\bibinfo {volume} {5}},\ \bibinfo
  {pages} {735} (\bibinfo {year} {2022})}\BibitemShut {NoStop}%
\bibitem [{\citenamefont {Reimers}\ \emph {et~al.}(2024)\citenamefont
  {Reimers}, \citenamefont {Odenbreit}, \citenamefont {\ifmmode~\check{S}\else
  \v{S}\fi{}mejkal}, \citenamefont {Strocov}, \citenamefont {Constantinou},
  \citenamefont {Hellenes}, \citenamefont {Jaeschke~Ubiergo}, \citenamefont
  {Campos}, \citenamefont {Bharadwaj}, \citenamefont {Chakraborty},
  \citenamefont {Denneulin}, \citenamefont {Shi}, \citenamefont
  {Dunin-Borkowski}, \citenamefont {Das}, \citenamefont {Kl\"{a}ui},
  \citenamefont {Sinova},\ and\ \citenamefont {Jourdan}}]{Reimers2023}%
  \BibitemOpen
  \bibfield  {author} {\bibinfo {author} {\bibfnamefont {S.}~\bibnamefont
  {Reimers}}, \bibinfo {author} {\bibfnamefont {L.}~\bibnamefont {Odenbreit}},
  \bibinfo {author} {\bibfnamefont {L.}~\bibnamefont {\ifmmode~\check{S}\else
  \v{S}\fi{}mejkal}}, \bibinfo {author} {\bibfnamefont {V.~N.}\ \bibnamefont
  {Strocov}}, \bibinfo {author} {\bibfnamefont {P.}~\bibnamefont
  {Constantinou}}, \bibinfo {author} {\bibfnamefont {A.~B.}\ \bibnamefont
  {Hellenes}}, \bibinfo {author} {\bibfnamefont {R.}~\bibnamefont
  {Jaeschke~Ubiergo}}, \bibinfo {author} {\bibfnamefont {W.~H.}\ \bibnamefont
  {Campos}}, \bibinfo {author} {\bibfnamefont {V.~K.}\ \bibnamefont
  {Bharadwaj}}, \bibinfo {author} {\bibfnamefont {A.}~\bibnamefont
  {Chakraborty}}, \bibinfo {author} {\bibfnamefont {T.}~\bibnamefont
  {Denneulin}}, \bibinfo {author} {\bibfnamefont {W.}~\bibnamefont {Shi}},
  \bibinfo {author} {\bibfnamefont {R.~E.}\ \bibnamefont {Dunin-Borkowski}},
  \bibinfo {author} {\bibfnamefont {S.}~\bibnamefont {Das}}, \bibinfo {author}
  {\bibfnamefont {M.}~\bibnamefont {Kl\"{a}ui}}, \bibinfo {author}
  {\bibfnamefont {J.}~\bibnamefont {Sinova}},\ and\ \bibinfo {author}
  {\bibfnamefont {M.}~\bibnamefont {Jourdan}},\ }\bibfield  {title} {\bibinfo
  {title} {Direct observation of altermagnetic band splitting in crsb thin
  films},\ }\href {https://doi.org/10.1038/s41467-024-46476-5} {\bibfield
  {journal} {\bibinfo  {journal} {Nat. Commun.}\ }\textbf {\bibinfo {volume}
  {15}},\ \bibinfo {pages} {2116} (\bibinfo {year} {2024})}\BibitemShut
  {NoStop}%
\bibitem [{\citenamefont {Lovesey}\ \emph {et~al.}(2023)\citenamefont
  {Lovesey}, \citenamefont {Khalyavin},\ and\ \citenamefont {van~der
  Laan}}]{Lovesey2023}%
  \BibitemOpen
  \bibfield  {author} {\bibinfo {author} {\bibfnamefont {S.~W.}\ \bibnamefont
  {Lovesey}}, \bibinfo {author} {\bibfnamefont {D.~D.}\ \bibnamefont
  {Khalyavin}},\ and\ \bibinfo {author} {\bibfnamefont {G.}~\bibnamefont
  {van~der Laan}},\ }\bibfield  {title} {\bibinfo {title} {{Templates for
  magnetic symmetry and altermagnetism in hexagonal MnTe}},\ }\href
  {https://doi.org/10.1103/PhysRevB.108.174437} {\bibfield  {journal} {\bibinfo
   {journal} {Phys. Rev. B}\ }\textbf {\bibinfo {volume} {108}},\ \bibinfo
  {pages} {174437} (\bibinfo {year} {2023})}\BibitemShut {NoStop}%
\bibitem [{\citenamefont {Mazin}\ \emph {et~al.}()\citenamefont {Mazin},
  \citenamefont {González-Hernández},\ and\ \citenamefont
  {Šmejkal}}]{Igor2023}%
  \BibitemOpen
  \bibfield  {author} {\bibinfo {author} {\bibfnamefont {I.}~\bibnamefont
  {Mazin}}, \bibinfo {author} {\bibfnamefont {R.}~\bibnamefont
  {González-Hernández}},\ and\ \bibinfo {author} {\bibfnamefont
  {L.}~\bibnamefont {Šmejkal}},\ }\href@noop {} {\bibinfo {title} {{Induced
  Monolayer Altermagnetism in MnP(S,Se)$_3$ and FeSe}}},\ \Eprint
  {https://arxiv.org/abs/arXiv:2309.02355} {arXiv:2309.02355} \BibitemShut
  {NoStop}%
\bibitem [{\citenamefont {Fedchenko}\ \emph {et~al.}(2024)\citenamefont
  {Fedchenko}, \citenamefont {Min\'ar}, \citenamefont {Akashdeep},
  \citenamefont {D'Souza}, \citenamefont {Vasilyev}, \citenamefont {Tkach},
  \citenamefont {Odenbreit}, \citenamefont {Nguyen}, \citenamefont
  {Kutnyakhov}, \citenamefont {Wind}, \citenamefont {Wenthaus}, \citenamefont
  {Scholz}, \citenamefont {Rossnagel}, \citenamefont {Hoesch}, \citenamefont
  {Aeschlimann}, \citenamefont {Stadtm\"{u}ller}, \citenamefont {Kl\"{a}ui},
  \citenamefont {Sch\"{o}nhense}, \citenamefont {Jungwirth}, \citenamefont
  {Hellenes}, \citenamefont {Jakob}, \citenamefont {\ifmmode~\check{S}\else
  \v{S}\fi{}mejkal}, \citenamefont {Sinova},\ and\ \citenamefont
  {Elmers}}]{Fedchenko2023}%
  \BibitemOpen
  \bibfield  {author} {\bibinfo {author} {\bibfnamefont {O.}~\bibnamefont
  {Fedchenko}}, \bibinfo {author} {\bibfnamefont {J.}~\bibnamefont {Min\'ar}},
  \bibinfo {author} {\bibfnamefont {A.}~\bibnamefont {Akashdeep}}, \bibinfo
  {author} {\bibfnamefont {S.~W.}\ \bibnamefont {D'Souza}}, \bibinfo {author}
  {\bibfnamefont {D.}~\bibnamefont {Vasilyev}}, \bibinfo {author}
  {\bibfnamefont {O.}~\bibnamefont {Tkach}}, \bibinfo {author} {\bibfnamefont
  {L.}~\bibnamefont {Odenbreit}}, \bibinfo {author} {\bibfnamefont
  {Q.}~\bibnamefont {Nguyen}}, \bibinfo {author} {\bibfnamefont
  {D.}~\bibnamefont {Kutnyakhov}}, \bibinfo {author} {\bibfnamefont
  {N.}~\bibnamefont {Wind}}, \bibinfo {author} {\bibfnamefont {L.}~\bibnamefont
  {Wenthaus}}, \bibinfo {author} {\bibfnamefont {M.}~\bibnamefont {Scholz}},
  \bibinfo {author} {\bibfnamefont {K.}~\bibnamefont {Rossnagel}}, \bibinfo
  {author} {\bibfnamefont {M.}~\bibnamefont {Hoesch}}, \bibinfo {author}
  {\bibfnamefont {M.}~\bibnamefont {Aeschlimann}}, \bibinfo {author}
  {\bibfnamefont {B.}~\bibnamefont {Stadtm\"{u}ller}}, \bibinfo {author}
  {\bibfnamefont {M.}~\bibnamefont {Kl\"{a}ui}}, \bibinfo {author}
  {\bibfnamefont {G.}~\bibnamefont {Sch\"{o}nhense}}, \bibinfo {author}
  {\bibfnamefont {T.}~\bibnamefont {Jungwirth}}, \bibinfo {author}
  {\bibfnamefont {A.~B.}\ \bibnamefont {Hellenes}}, \bibinfo {author}
  {\bibfnamefont {G.}~\bibnamefont {Jakob}}, \bibinfo {author} {\bibfnamefont
  {L.}~\bibnamefont {\ifmmode~\check{S}\else \v{S}\fi{}mejkal}}, \bibinfo
  {author} {\bibfnamefont {J.}~\bibnamefont {Sinova}},\ and\ \bibinfo {author}
  {\bibfnamefont {H.-J.}\ \bibnamefont {Elmers}},\ }\bibfield  {title}
  {\bibinfo {title} {{Observation of time-reversal symmetry breaking in the
  band structure of altermagnetic RuO$_2$}},\ }\href
  {https://doi.org/10.1126/sciadv.adj4883} {\bibfield  {journal} {\bibinfo
  {journal} {Sci. Adv.}\ }\textbf {\bibinfo {volume} {10}},\ \bibinfo {pages}
  {eadj4883} (\bibinfo {year} {2024})}\BibitemShut {NoStop}%
\bibitem [{\citenamefont {Bai}\ \emph {et~al.}(2023)\citenamefont {Bai},
  \citenamefont {Zhang}, \citenamefont {Zhou}, \citenamefont {Chen},
  \citenamefont {Wan}, \citenamefont {Han}, \citenamefont {Zhu}, \citenamefont
  {Liang}, \citenamefont {Su}, \citenamefont {Han}, \citenamefont {Pan},\ and\
  \citenamefont {Song}}]{Bai2023}%
  \BibitemOpen
  \bibfield  {author} {\bibinfo {author} {\bibfnamefont {H.}~\bibnamefont
  {Bai}}, \bibinfo {author} {\bibfnamefont {Y.~C.}\ \bibnamefont {Zhang}},
  \bibinfo {author} {\bibfnamefont {Y.~J.}\ \bibnamefont {Zhou}}, \bibinfo
  {author} {\bibfnamefont {P.}~\bibnamefont {Chen}}, \bibinfo {author}
  {\bibfnamefont {C.~H.}\ \bibnamefont {Wan}}, \bibinfo {author} {\bibfnamefont
  {L.}~\bibnamefont {Han}}, \bibinfo {author} {\bibfnamefont {W.~X.}\
  \bibnamefont {Zhu}}, \bibinfo {author} {\bibfnamefont {S.~X.}\ \bibnamefont
  {Liang}}, \bibinfo {author} {\bibfnamefont {Y.~C.}\ \bibnamefont {Su}},
  \bibinfo {author} {\bibfnamefont {X.~F.}\ \bibnamefont {Han}}, \bibinfo
  {author} {\bibfnamefont {F.}~\bibnamefont {Pan}},\ and\ \bibinfo {author}
  {\bibfnamefont {C.}~\bibnamefont {Song}},\ }\bibfield  {title} {\bibinfo
  {title} {{Efficient Spin-to-Charge Conversion via Altermagnetic Spin
  Splitting Effect in Antiferromagnet ${\mathrm{RuO}}_{2}$}},\ }\href
  {https://doi.org/10.1103/PhysRevLett.130.216701} {\bibfield  {journal}
  {\bibinfo  {journal} {Phys. Rev. Lett.}\ }\textbf {\bibinfo {volume} {130}},\
  \bibinfo {pages} {216701} (\bibinfo {year} {2023})}\BibitemShut {NoStop}%
\bibitem [{\citenamefont {Grzybowski}\ \emph {et~al.}(2024)\citenamefont
  {Grzybowski}, \citenamefont {Autieri}, \citenamefont {Domagala},
  \citenamefont {Krasucki}, \citenamefont {Kaleta}, \citenamefont {Kret},
  \citenamefont {Gas}, \citenamefont {Sawicki}, \citenamefont {Bo$\ddot{\rm
  z}$ek}, \citenamefont {Suffczy\'nski},\ and\ \citenamefont
  {Pacuski}}]{Grzybowski2023}%
  \BibitemOpen
  \bibfield  {author} {\bibinfo {author} {\bibfnamefont {M.~J.}\ \bibnamefont
  {Grzybowski}}, \bibinfo {author} {\bibfnamefont {C.}~\bibnamefont {Autieri}},
  \bibinfo {author} {\bibfnamefont {J.}~\bibnamefont {Domagala}}, \bibinfo
  {author} {\bibfnamefont {C.}~\bibnamefont {Krasucki}}, \bibinfo {author}
  {\bibfnamefont {A.}~\bibnamefont {Kaleta}}, \bibinfo {author} {\bibfnamefont
  {S.}~\bibnamefont {Kret}}, \bibinfo {author} {\bibfnamefont {K.}~\bibnamefont
  {Gas}}, \bibinfo {author} {\bibfnamefont {M.}~\bibnamefont {Sawicki}},
  \bibinfo {author} {\bibfnamefont {R.}~\bibnamefont {Bo$\ddot{\rm z}$ek}},
  \bibinfo {author} {\bibfnamefont {J.}~\bibnamefont {Suffczy\'nski}},\ and\
  \bibinfo {author} {\bibfnamefont {W.}~\bibnamefont {Pacuski}},\ }\bibfield
  {title} {\bibinfo {title} {{Wurtzite vs. rock-salt MnSe epitaxy: electronic
  and altermagnetic properties}},\ }\href {https://doi.org/10.1039/d3nr04798a}
  {\bibfield  {journal} {\bibinfo  {journal} {Nanoscale}\ }\textbf {\bibinfo
  {volume} {16}},\ \bibinfo {pages} {6259} (\bibinfo {year}
  {2024})}\BibitemShut {NoStop}%
\bibitem [{\citenamefont {Kluczyk}\ \emph {et~al.}(2024)\citenamefont
  {Kluczyk}, \citenamefont {Gas}, \citenamefont {Grzybowski}, \citenamefont
  {Skupinski}, \citenamefont {Borysiewicz}, \citenamefont {Fas}, \citenamefont
  {Suffczynski}, \citenamefont {Domagala}, \citenamefont {Grasza},
  \citenamefont {Mycielski}, \citenamefont {Baj}, \citenamefont {Ahn},
  \citenamefont {Vyborny}, \citenamefont {Sawicki},\ and\ \citenamefont
  {Gryglas-Boryseiwicz}}]{Kluczyk2023}%
  \BibitemOpen
  \bibfield  {author} {\bibinfo {author} {\bibfnamefont {K.~P.}\ \bibnamefont
  {Kluczyk}}, \bibinfo {author} {\bibfnamefont {K.}~\bibnamefont {Gas}},
  \bibinfo {author} {\bibfnamefont {M.~J.}\ \bibnamefont {Grzybowski}},
  \bibinfo {author} {\bibfnamefont {P.}~\bibnamefont {Skupinski}}, \bibinfo
  {author} {\bibfnamefont {M.~A.}\ \bibnamefont {Borysiewicz}}, \bibinfo
  {author} {\bibfnamefont {T.}~\bibnamefont {Fas}}, \bibinfo {author}
  {\bibfnamefont {J.}~\bibnamefont {Suffczynski}}, \bibinfo {author}
  {\bibfnamefont {J.~Z.}\ \bibnamefont {Domagala}}, \bibinfo {author}
  {\bibfnamefont {K.}~\bibnamefont {Grasza}}, \bibinfo {author} {\bibfnamefont
  {A.}~\bibnamefont {Mycielski}}, \bibinfo {author} {\bibfnamefont
  {M.}~\bibnamefont {Baj}}, \bibinfo {author} {\bibfnamefont {K.~H.}\
  \bibnamefont {Ahn}}, \bibinfo {author} {\bibfnamefont {K.}~\bibnamefont
  {Vyborny}}, \bibinfo {author} {\bibfnamefont {M.}~\bibnamefont {Sawicki}},\
  and\ \bibinfo {author} {\bibfnamefont {M.}~\bibnamefont
  {Gryglas-Boryseiwicz}},\ }\bibfield  {title} {\bibinfo {title} {{Coexistence
  of anomalous Hall effect and weak magnetization in a nominally collinear
  antiferromagnet MnTe}},\ }\href {https://doi.org/10.1103/PhysRevB.110.155201}
  {\bibfield  {journal} {\bibinfo  {journal} {Phys. Rev. B}\ }\textbf {\bibinfo
  {volume} {110}},\ \bibinfo {pages} {155201} (\bibinfo {year}
  {2024})}\BibitemShut {NoStop}%
\bibitem [{\citenamefont {Aoyama}\ and\ \citenamefont
  {Ohgushi}(2024)}]{Takuya2023}%
  \BibitemOpen
  \bibfield  {author} {\bibinfo {author} {\bibfnamefont {T.}~\bibnamefont
  {Aoyama}}\ and\ \bibinfo {author} {\bibfnamefont {K.}~\bibnamefont
  {Ohgushi}},\ }\bibfield  {title} {\bibinfo {title} {{Piezomagnetic properties
  in altermagnetic MnTe}},\ }\href
  {https://doi.org/10.1103/PhysRevMaterials.8.L041402} {\bibfield  {journal}
  {\bibinfo  {journal} {Phys. Rev. Mater.}\ }\textbf {\bibinfo {volume} {8}},\
  \bibinfo {pages} {L041402} (\bibinfo {year} {2024})}\BibitemShut {NoStop}%
\bibitem [{\citenamefont {Cuono}\ \emph {et~al.}(2023)\citenamefont {Cuono},
  \citenamefont {Sattigeri}, \citenamefont {Skolimowski},\ and\ \citenamefont
  {Autieri}}]{Cuono2023}%
  \BibitemOpen
  \bibfield  {author} {\bibinfo {author} {\bibfnamefont {G.}~\bibnamefont
  {Cuono}}, \bibinfo {author} {\bibfnamefont {R.~M.}\ \bibnamefont
  {Sattigeri}}, \bibinfo {author} {\bibfnamefont {J.}~\bibnamefont
  {Skolimowski}},\ and\ \bibinfo {author} {\bibfnamefont {C.}~\bibnamefont
  {Autieri}},\ }\bibfield  {title} {\bibinfo {title} {{Orbital-selective
  altermagnetism and correlation-enhanced spin-splitting in strongly-correlated
  transition metal oxides}},\ }\href
  {https://doi.org/10.1016/j.jmmm.2023.171163} {\bibfield  {journal} {\bibinfo
  {journal} {J. Magn. Magn. Mat.}\ }\textbf {\bibinfo {volume} {586}},\
  \bibinfo {pages} {171163} (\bibinfo {year} {2023})}\BibitemShut {NoStop}%
\bibitem [{\citenamefont {Lin}\ \emph {et~al.}()\citenamefont {Lin},
  \citenamefont {Chen}, \citenamefont {Lu}, \citenamefont {Liang},
  \citenamefont {Feng}, \citenamefont {Yamagami}, \citenamefont {Osiecki},
  \citenamefont {Leandersson}, \citenamefont {Thiagarajan}, \citenamefont
  {Liu}, \citenamefont {Felser},\ and\ \citenamefont {Ma}}]{Lin2024}%
  \BibitemOpen
  \bibfield  {author} {\bibinfo {author} {\bibfnamefont {Z.}~\bibnamefont
  {Lin}}, \bibinfo {author} {\bibfnamefont {D.}~\bibnamefont {Chen}}, \bibinfo
  {author} {\bibfnamefont {W.}~\bibnamefont {Lu}}, \bibinfo {author}
  {\bibfnamefont {X.}~\bibnamefont {Liang}}, \bibinfo {author} {\bibfnamefont
  {S.}~\bibnamefont {Feng}}, \bibinfo {author} {\bibfnamefont {K.}~\bibnamefont
  {Yamagami}}, \bibinfo {author} {\bibfnamefont {J.}~\bibnamefont {Osiecki}},
  \bibinfo {author} {\bibfnamefont {M.}~\bibnamefont {Leandersson}}, \bibinfo
  {author} {\bibfnamefont {B.}~\bibnamefont {Thiagarajan}}, \bibinfo {author}
  {\bibfnamefont {J.}~\bibnamefont {Liu}}, \bibinfo {author} {\bibfnamefont
  {C.}~\bibnamefont {Felser}},\ and\ \bibinfo {author} {\bibfnamefont
  {J.}~\bibnamefont {Ma}},\ }\href@noop {} {\bibinfo {title} {{Observation of
  Giant Spin Splitting and $d$-wave Spin Texture in Room Temperature
  Altermagnet RuO$_2$}}},\ \Eprint {https://arxiv.org/abs/arXiv:2402.04995}
  {arXiv:2402.04995} \BibitemShut {NoStop}%
\bibitem [{\citenamefont {Lee}\ \emph {et~al.}(2024)\citenamefont {Lee},
  \citenamefont {Lee}, \citenamefont {Jung}, \citenamefont {Jung},
  \citenamefont {Kim}, \citenamefont {Lee}, \citenamefont {Seok}, \citenamefont
  {Kim}, \citenamefont {Park}, \citenamefont {\ifmmode~\check{S}\else
  \v{S}\fi{}mejkal}, \citenamefont {Kang},\ and\ \citenamefont
  {Kim}}]{Lee2024}%
  \BibitemOpen
  \bibfield  {author} {\bibinfo {author} {\bibfnamefont {S.}~\bibnamefont
  {Lee}}, \bibinfo {author} {\bibfnamefont {S.}~\bibnamefont {Lee}}, \bibinfo
  {author} {\bibfnamefont {S.}~\bibnamefont {Jung}}, \bibinfo {author}
  {\bibfnamefont {J.}~\bibnamefont {Jung}}, \bibinfo {author} {\bibfnamefont
  {D.}~\bibnamefont {Kim}}, \bibinfo {author} {\bibfnamefont {Y.}~\bibnamefont
  {Lee}}, \bibinfo {author} {\bibfnamefont {B.}~\bibnamefont {Seok}}, \bibinfo
  {author} {\bibfnamefont {J.}~\bibnamefont {Kim}}, \bibinfo {author}
  {\bibfnamefont {B.~G.}\ \bibnamefont {Park}}, \bibinfo {author}
  {\bibfnamefont {L.}~\bibnamefont {\ifmmode~\check{S}\else \v{S}\fi{}mejkal}},
  \bibinfo {author} {\bibfnamefont {C.-J.}\ \bibnamefont {Kang}},\ and\
  \bibinfo {author} {\bibfnamefont {C.}~\bibnamefont {Kim}},\ }\bibfield
  {title} {\bibinfo {title} {{Broken Kramers Degeneracy in Altermagnetic
  MnTe}},\ }\href {https://doi.org/10.1103/PhysRevLett.132.036702} {\bibfield
  {journal} {\bibinfo  {journal} {Phys. Rev. Lett.}\ }\textbf {\bibinfo
  {volume} {132}},\ \bibinfo {pages} {036702} (\bibinfo {year}
  {2024})}\BibitemShut {NoStop}%
\bibitem [{\citenamefont {Guo}\ \emph {et~al.}(2023)\citenamefont {Guo},
  \citenamefont {Liu}, \citenamefont {Janson}, \citenamefont {Fulga},
  \citenamefont {van~den Brink},\ and\ \citenamefont {Facio}}]{Guo2023}%
  \BibitemOpen
  \bibfield  {author} {\bibinfo {author} {\bibfnamefont {Y.}~\bibnamefont
  {Guo}}, \bibinfo {author} {\bibfnamefont {H.}~\bibnamefont {Liu}}, \bibinfo
  {author} {\bibfnamefont {O.}~\bibnamefont {Janson}}, \bibinfo {author}
  {\bibfnamefont {I.~C.}\ \bibnamefont {Fulga}}, \bibinfo {author}
  {\bibfnamefont {J.}~\bibnamefont {van~den Brink}},\ and\ \bibinfo {author}
  {\bibfnamefont {J.~I.}\ \bibnamefont {Facio}},\ }\bibfield  {title} {\bibinfo
  {title} {{Spin-split collinear antiferromagnets: A large-scale ab-initio
  study}},\ }\href {https://doi.org/10.1016/j.mtphys.2023.100991} {\bibfield
  {journal} {\bibinfo  {journal} {Mater. Today Phys.}\ }\textbf {\bibinfo
  {volume} {32}},\ \bibinfo {pages} {100991} (\bibinfo {year}
  {2023})}\BibitemShut {NoStop}%
\bibitem [{\citenamefont {Volovik}(2009)}]{Volovik2009}%
  \BibitemOpen
  \bibfield  {author} {\bibinfo {author} {\bibfnamefont {G.~E.}\ \bibnamefont
  {Volovik}},\ }\href
  {https://doi.org/10.1093/acprof:oso/9780199564842.001.0001} {\emph {\bibinfo
  {title} {{The Universe in a Helium Droplet}}}}\ (\bibinfo  {publisher}
  {Oxford University Press, Oxford, UK},\ \bibinfo {year} {2009})\BibitemShut
  {NoStop}%
\bibitem [{\citenamefont {Fu}\ and\ \citenamefont {Berg}(2010)}]{FuBerg2010}%
  \BibitemOpen
  \bibfield  {author} {\bibinfo {author} {\bibfnamefont {L.}~\bibnamefont
  {Fu}}\ and\ \bibinfo {author} {\bibfnamefont {E.}~\bibnamefont {Berg}},\
  }\bibfield  {title} {\bibinfo {title} {{Odd-Parity Topological
  Superconductors: Theory and Application to
  ${\mathrm{Cu}}_{x}{\mathrm{Bi}}_{2}{\mathrm{Se}}_{3}$}},\ }\href
  {https://doi.org/10.1103/PhysRevLett.105.097001} {\bibfield  {journal}
  {\bibinfo  {journal} {Phys. Rev. Lett.}\ }\textbf {\bibinfo {volume} {105}},\
  \bibinfo {pages} {097001} (\bibinfo {year} {2010})}\BibitemShut {NoStop}%
\bibitem [{\citenamefont {Fu}(2014)}]{LiangFu2014}%
  \BibitemOpen
  \bibfield  {author} {\bibinfo {author} {\bibfnamefont {L.}~\bibnamefont
  {Fu}},\ }\bibfield  {title} {\bibinfo {title} {{Odd-parity topological
  superconductor with nematic order: Application to
  ${\mathrm{Cu}}_{x}{\mathrm{Bi}}_{2}{\mathrm{Se}}_{3}$}},\ }\href
  {https://doi.org/10.1103/PhysRevB.90.100509} {\bibfield  {journal} {\bibinfo
  {journal} {Phys. Rev. B}\ }\textbf {\bibinfo {volume} {90}},\ \bibinfo
  {pages} {100509} (\bibinfo {year} {2014})}\BibitemShut {NoStop}%
\bibitem [{\citenamefont {Roy}\ \emph {et~al.}(2017)\citenamefont {Roy},
  \citenamefont {Alavirad},\ and\ \citenamefont {Sau}}]{Alavirad2017}%
  \BibitemOpen
  \bibfield  {author} {\bibinfo {author} {\bibfnamefont {B.}~\bibnamefont
  {Roy}}, \bibinfo {author} {\bibfnamefont {Y.}~\bibnamefont {Alavirad}},\ and\
  \bibinfo {author} {\bibfnamefont {J.~D.}\ \bibnamefont {Sau}},\ }\bibfield
  {title} {\bibinfo {title} {{Global Phase Diagram of a Three-Dimensional Dirty
  Topological Superconductor}},\ }\href
  {https://doi.org/10.1103/PhysRevLett.118.227002} {\bibfield  {journal}
  {\bibinfo  {journal} {Phys. Rev. Lett.}\ }\textbf {\bibinfo {volume} {118}},\
  \bibinfo {pages} {227002} (\bibinfo {year} {2017})}\BibitemShut {NoStop}%
\bibitem [{\citenamefont {Roy}\ \emph {et~al.}(2019)\citenamefont {Roy},
  \citenamefont {Ghorashi}, \citenamefont {Foster},\ and\ \citenamefont
  {Nevidomskyy}}]{RoyFosterNevidomskyy2019}%
  \BibitemOpen
  \bibfield  {author} {\bibinfo {author} {\bibfnamefont {B.}~\bibnamefont
  {Roy}}, \bibinfo {author} {\bibfnamefont {S.~A.~A.}\ \bibnamefont
  {Ghorashi}}, \bibinfo {author} {\bibfnamefont {M.~S.}\ \bibnamefont
  {Foster}},\ and\ \bibinfo {author} {\bibfnamefont {A.~H.}\ \bibnamefont
  {Nevidomskyy}},\ }\bibfield  {title} {\bibinfo {title} {{Topological
  superconductivity of spin-$3/2$ carriers in a three-dimensional doped
  Luttinger semimetal}},\ }\href {https://doi.org/10.1103/PhysRevB.99.054505}
  {\bibfield  {journal} {\bibinfo  {journal} {Phys. Rev. B}\ }\textbf {\bibinfo
  {volume} {99}},\ \bibinfo {pages} {054505} (\bibinfo {year}
  {2019})}\BibitemShut {NoStop}%
\bibitem [{\citenamefont {Kriener}\ \emph {et~al.}(2011)\citenamefont
  {Kriener}, \citenamefont {Segawa}, \citenamefont {Ren}, \citenamefont
  {Sasaki},\ and\ \citenamefont {Ando}}]{Ando2011PRL}%
  \BibitemOpen
  \bibfield  {author} {\bibinfo {author} {\bibfnamefont {M.}~\bibnamefont
  {Kriener}}, \bibinfo {author} {\bibfnamefont {K.}~\bibnamefont {Segawa}},
  \bibinfo {author} {\bibfnamefont {Z.}~\bibnamefont {Ren}}, \bibinfo {author}
  {\bibfnamefont {S.}~\bibnamefont {Sasaki}},\ and\ \bibinfo {author}
  {\bibfnamefont {Y.}~\bibnamefont {Ando}},\ }\bibfield  {title} {\bibinfo
  {title} {{Bulk Superconducting Phase with a Full Energy Gap in the Doped
  Topological Insulator
  ${\mathrm{Cu}}_{x}{\mathrm{Bi}}_{2}{\mathrm{Se}}_{3}$}},\ }\href
  {https://doi.org/10.1103/PhysRevLett.106.127004} {\bibfield  {journal}
  {\bibinfo  {journal} {Phys. Rev. Lett.}\ }\textbf {\bibinfo {volume} {106}},\
  \bibinfo {pages} {127004} (\bibinfo {year} {2011})}\BibitemShut {NoStop}%
\bibitem [{\citenamefont {Sasaki}\ \emph {et~al.}(2012)\citenamefont {Sasaki},
  \citenamefont {Ren}, \citenamefont {Taskin}, \citenamefont {Segawa},
  \citenamefont {Fu},\ and\ \citenamefont {Ando}}]{Ando2012PRL}%
  \BibitemOpen
  \bibfield  {author} {\bibinfo {author} {\bibfnamefont {S.}~\bibnamefont
  {Sasaki}}, \bibinfo {author} {\bibfnamefont {Z.}~\bibnamefont {Ren}},
  \bibinfo {author} {\bibfnamefont {A.~A.}\ \bibnamefont {Taskin}}, \bibinfo
  {author} {\bibfnamefont {K.}~\bibnamefont {Segawa}}, \bibinfo {author}
  {\bibfnamefont {L.}~\bibnamefont {Fu}},\ and\ \bibinfo {author}
  {\bibfnamefont {Y.}~\bibnamefont {Ando}},\ }\bibfield  {title} {\bibinfo
  {title} {{Odd-Parity Pairing and Topological Superconductivity in a Strongly
  Spin-Orbit Coupled Semiconductor}},\ }\href
  {https://doi.org/10.1103/PhysRevLett.109.217004} {\bibfield  {journal}
  {\bibinfo  {journal} {Phys. Rev. Lett.}\ }\textbf {\bibinfo {volume} {109}},\
  \bibinfo {pages} {217004} (\bibinfo {year} {2012})}\BibitemShut {NoStop}%
\bibitem [{\citenamefont {Novak}\ \emph {et~al.}(2013)\citenamefont {Novak},
  \citenamefont {Sasaki}, \citenamefont {Kriener}, \citenamefont {Segawa},\
  and\ \citenamefont {Ando}}]{Ando2013PRB}%
  \BibitemOpen
  \bibfield  {author} {\bibinfo {author} {\bibfnamefont {M.}~\bibnamefont
  {Novak}}, \bibinfo {author} {\bibfnamefont {S.}~\bibnamefont {Sasaki}},
  \bibinfo {author} {\bibfnamefont {M.}~\bibnamefont {Kriener}}, \bibinfo
  {author} {\bibfnamefont {K.}~\bibnamefont {Segawa}},\ and\ \bibinfo {author}
  {\bibfnamefont {Y.}~\bibnamefont {Ando}},\ }\bibfield  {title} {\bibinfo
  {title} {{Unusual nature of fully gapped superconductivity in In-doped
  SnTe}},\ }\href {https://doi.org/10.1103/PhysRevB.88.140502} {\bibfield
  {journal} {\bibinfo  {journal} {Phys. Rev. B}\ }\textbf {\bibinfo {volume}
  {88}},\ \bibinfo {pages} {140502} (\bibinfo {year} {2013})}\BibitemShut
  {NoStop}%
\bibitem [{\citenamefont {Ghorashi}\ \emph {et~al.}(2019)\citenamefont
  {Ghorashi}, \citenamefont {Hu}, \citenamefont {Hughes},\ and\ \citenamefont
  {Rossi}}]{Ghorashi2019}%
  \BibitemOpen
  \bibfield  {author} {\bibinfo {author} {\bibfnamefont {S.~A.~A.}\
  \bibnamefont {Ghorashi}}, \bibinfo {author} {\bibfnamefont {X.}~\bibnamefont
  {Hu}}, \bibinfo {author} {\bibfnamefont {T.~L.}\ \bibnamefont {Hughes}},\
  and\ \bibinfo {author} {\bibfnamefont {E.}~\bibnamefont {Rossi}},\ }\bibfield
   {title} {\bibinfo {title} {{Second-order Dirac superconductors and magnetic
  field induced Majorana hinge modes}},\ }\href
  {https://doi.org/10.1103/PhysRevB.100.020509} {\bibfield  {journal} {\bibinfo
   {journal} {Phys. Rev. B}\ }\textbf {\bibinfo {volume} {100}},\ \bibinfo
  {pages} {020509} (\bibinfo {year} {2019})}\BibitemShut {NoStop}%
\bibitem [{\citenamefont {Roy}(2020)}]{BRoy2020}%
  \BibitemOpen
  \bibfield  {author} {\bibinfo {author} {\bibfnamefont {B.}~\bibnamefont
  {Roy}},\ }\bibfield  {title} {\bibinfo {title} {{Higher-order topological
  superconductors in $\mathcal{P}$-, $\mathcal{T}$-odd quadrupolar Dirac
  materials}},\ }\href {https://doi.org/10.1103/PhysRevB.101.220506} {\bibfield
   {journal} {\bibinfo  {journal} {Phys. Rev. B}\ }\textbf {\bibinfo {volume}
  {101}},\ \bibinfo {pages} {220506} (\bibinfo {year} {2020})}\BibitemShut
  {NoStop}%
\bibitem [{\citenamefont {Roy}\ and\ \citenamefont {Juri\ifmmode \check{c}\else
  \v{c}\fi{}i\ifmmode~\acute{c}\else \'{c}\fi{}}(2021)}]{RoyJuricic2021}%
  \BibitemOpen
  \bibfield  {author} {\bibinfo {author} {\bibfnamefont {B.}~\bibnamefont
  {Roy}}\ and\ \bibinfo {author} {\bibfnamefont {V.}~\bibnamefont {Juri\ifmmode
  \check{c}\else \v{c}\fi{}i\ifmmode~\acute{c}\else \'{c}\fi{}}},\ }\bibfield
  {title} {\bibinfo {title} {{Mixed-parity octupolar pairing and corner
  Majorana modes in three dimensions}},\ }\href
  {https://doi.org/10.1103/PhysRevB.104.L180503} {\bibfield  {journal}
  {\bibinfo  {journal} {Phys. Rev. B}\ }\textbf {\bibinfo {volume} {104}},\
  \bibinfo {pages} {L180503} (\bibinfo {year} {2021})}\BibitemShut {NoStop}%
\bibitem [{\citenamefont {Castro~Neto}\ \emph {et~al.}(2009)\citenamefont
  {Castro~Neto}, \citenamefont {Guinea}, \citenamefont {Peres}, \citenamefont
  {Novoselov},\ and\ \citenamefont {Geim}}]{GrapheneRMP2009}%
  \BibitemOpen
  \bibfield  {author} {\bibinfo {author} {\bibfnamefont {A.~H.}\ \bibnamefont
  {Castro~Neto}}, \bibinfo {author} {\bibfnamefont {F.}~\bibnamefont {Guinea}},
  \bibinfo {author} {\bibfnamefont {N.~M.~R.}\ \bibnamefont {Peres}}, \bibinfo
  {author} {\bibfnamefont {K.~S.}\ \bibnamefont {Novoselov}},\ and\ \bibinfo
  {author} {\bibfnamefont {A.~K.}\ \bibnamefont {Geim}},\ }\bibfield  {title}
  {\bibinfo {title} {{The electronic properties of graphene}},\ }\href
  {https://doi.org/10.1103/RevModPhys.81.109} {\bibfield  {journal} {\bibinfo
  {journal} {Rev. Mod. Phys.}\ }\textbf {\bibinfo {volume} {81}},\ \bibinfo
  {pages} {109} (\bibinfo {year} {2009})}\BibitemShut {NoStop}%
\bibitem [{SMa()}]{SMaltermagnet}%
  \BibitemOpen
  \href@noop {} {\bibinfo  {journal} {See Supplemental Material at XXX-XXXX for
  derivation of low-energy models from minimal tight-binding Hamiltonian,
  symmetry analysis of free fermion systems, energetics of spin nematic orders,
  lattice realization of charge, spin, paired nematic oders, solutions of
  mean-field gap equations, and the signatures of trigonal warping in BBLG and
  RTLG on altermagnets}\ }\BibitemShut {NoStop}%
\bibitem [{\citenamefont {Herbut}\ \emph {et~al.}(2009)\citenamefont {Herbut},
  \citenamefont {Juri\ifmmode \check{c}\else \v{c}\fi{}i\ifmmode~\acute{c}\else
  \'{c}\fi{}},\ and\ \citenamefont {Roy}}]{HJR2009}%
  \BibitemOpen
\bibfield  {journal} {  }\bibfield  {author} {\bibinfo {author} {\bibfnamefont
  {I.~F.}\ \bibnamefont {Herbut}}, \bibinfo {author} {\bibfnamefont
  {V.}~\bibnamefont {Juri\ifmmode \check{c}\else
  \v{c}\fi{}i\ifmmode~\acute{c}\else \'{c}\fi{}}},\ and\ \bibinfo {author}
  {\bibfnamefont {B.}~\bibnamefont {Roy}},\ }\bibfield  {title} {\bibinfo
  {title} {{Theory of interacting electrons on the honeycomb lattice}},\ }\href
  {https://doi.org/10.1103/PhysRevB.79.085116} {\bibfield  {journal} {\bibinfo
  {journal} {Phys. Rev. B}\ }\textbf {\bibinfo {volume} {79}},\ \bibinfo
  {pages} {085116} (\bibinfo {year} {2009})}\BibitemShut {NoStop}%
\bibitem [{\citenamefont {Zhang}\ \emph {et~al.}(2010)\citenamefont {Zhang},
  \citenamefont {Sahu}, \citenamefont {Min},\ and\ \citenamefont
  {MacDonald}}]{FanZhang2010}%
  \BibitemOpen
  \bibfield  {author} {\bibinfo {author} {\bibfnamefont {F.}~\bibnamefont
  {Zhang}}, \bibinfo {author} {\bibfnamefont {B.}~\bibnamefont {Sahu}},
  \bibinfo {author} {\bibfnamefont {H.}~\bibnamefont {Min}},\ and\ \bibinfo
  {author} {\bibfnamefont {A.~H.}\ \bibnamefont {MacDonald}},\ }\bibfield
  {title} {\bibinfo {title} {{Band structure of $ABC$-stacked graphene
  trilayers}},\ }\href {https://doi.org/10.1103/PhysRevB.82.035409} {\bibfield
  {journal} {\bibinfo  {journal} {Phys. Rev. B}\ }\textbf {\bibinfo {volume}
  {82}},\ \bibinfo {pages} {035409} (\bibinfo {year} {2010})}\BibitemShut
  {NoStop}%
\bibitem [{\citenamefont {Vafek}(2010)}]{Vafek2010}%
  \BibitemOpen
  \bibfield  {author} {\bibinfo {author} {\bibfnamefont {O.}~\bibnamefont
  {Vafek}},\ }\bibfield  {title} {\bibinfo {title} {{Interacting fermions on
  the honeycomb bilayer: From weak to strong coupling}},\ }\href
  {https://doi.org/10.1103/PhysRevB.82.205106} {\bibfield  {journal} {\bibinfo
  {journal} {Phys. Rev. B}\ }\textbf {\bibinfo {volume} {82}},\ \bibinfo
  {pages} {205106} (\bibinfo {year} {2010})}\BibitemShut {NoStop}%
\bibitem [{\citenamefont {Roy}(2013)}]{BR2013}%
  \BibitemOpen
  \bibfield  {author} {\bibinfo {author} {\bibfnamefont {B.}~\bibnamefont
  {Roy}},\ }\bibfield  {title} {\bibinfo {title} {{Classification of massive
  and gapless phases in bilayer graphene}},\ }\href
  {https://doi.org/10.1103/PhysRevB.88.075415} {\bibfield  {journal} {\bibinfo
  {journal} {Phys. Rev. B}\ }\textbf {\bibinfo {volume} {88}},\ \bibinfo
  {pages} {075415} (\bibinfo {year} {2013})}\BibitemShut {NoStop}%
\bibitem [{\citenamefont {Szab\'o}\ and\ \citenamefont
  {Roy}(2021)}]{Szabo2021}%
  \BibitemOpen
  \bibfield  {author} {\bibinfo {author} {\bibfnamefont {A.~L.}\ \bibnamefont
  {Szab\'o}}\ and\ \bibinfo {author} {\bibfnamefont {B.}~\bibnamefont {Roy}},\
  }\bibfield  {title} {\bibinfo {title} {{Extended Hubbard model in undoped and
  doped monolayer and bilayer graphene: Selection rules and organizing
  principle among competing orders}},\ }\href
  {https://doi.org/10.1103/PhysRevB.103.205135} {\bibfield  {journal} {\bibinfo
   {journal} {Phys. Rev. B}\ }\textbf {\bibinfo {volume} {103}},\ \bibinfo
  {pages} {205135} (\bibinfo {year} {2021})}\BibitemShut {NoStop}%
\bibitem [{\citenamefont {Szab\'o}\ and\ \citenamefont
  {Roy}(2022)}]{Szabo2022}%
  \BibitemOpen
  \bibfield  {author} {\bibinfo {author} {\bibfnamefont {A.~L.}\ \bibnamefont
  {Szab\'o}}\ and\ \bibinfo {author} {\bibfnamefont {B.}~\bibnamefont {Roy}},\
  }\bibfield  {title} {\bibinfo {title} {{Metals, fractional metals, and
  superconductivity in rhombohedral trilayer graphene}},\ }\href
  {https://doi.org/10.1103/PhysRevB.105.L081407} {\bibfield  {journal}
  {\bibinfo  {journal} {Phys. Rev. B}\ }\textbf {\bibinfo {volume} {105}},\
  \bibinfo {pages} {L081407} (\bibinfo {year} {2022})}\BibitemShut {NoStop}%
\bibitem [{\citenamefont {{Zhou}}\ \emph
  {et~al.}(2021{\natexlab{a}})\citenamefont {{Zhou}}, \citenamefont {{Xie}},
  \citenamefont {{Ghazaryan}}, \citenamefont {{Holder}}, \citenamefont
  {{Ehrets}}, \citenamefont {{Spanton}}, \citenamefont {{Taniguchi}},
  \citenamefont {{Watanabe}}, \citenamefont {{Berg}}, \citenamefont
  {{Serbyn}},\ and\ \citenamefont {{Young}}}]{Young2021I}%
  \BibitemOpen
  \bibfield  {author} {\bibinfo {author} {\bibfnamefont {H.}~\bibnamefont
  {{Zhou}}}, \bibinfo {author} {\bibfnamefont {T.}~\bibnamefont {{Xie}}},
  \bibinfo {author} {\bibfnamefont {A.}~\bibnamefont {{Ghazaryan}}}, \bibinfo
  {author} {\bibfnamefont {T.}~\bibnamefont {{Holder}}}, \bibinfo {author}
  {\bibfnamefont {J.~R.}\ \bibnamefont {{Ehrets}}}, \bibinfo {author}
  {\bibfnamefont {E.~M.}\ \bibnamefont {{Spanton}}}, \bibinfo {author}
  {\bibfnamefont {T.}~\bibnamefont {{Taniguchi}}}, \bibinfo {author}
  {\bibfnamefont {K.}~\bibnamefont {{Watanabe}}}, \bibinfo {author}
  {\bibfnamefont {E.}~\bibnamefont {{Berg}}}, \bibinfo {author} {\bibfnamefont
  {M.}~\bibnamefont {{Serbyn}}},\ and\ \bibinfo {author} {\bibfnamefont
  {A.~F.}\ \bibnamefont {{Young}}},\ }\bibfield  {title} {\bibinfo {title}
  {{Half- and quarter-metals in rhombohedral trilayer graphene}},\ }\href
  {https://doi.org/10.1038/s41586-021-03938-w} {\bibfield  {journal} {\bibinfo
  {journal} {\nat}\ }\textbf {\bibinfo {volume} {598}},\ \bibinfo {pages} {429}
  (\bibinfo {year} {2021}{\natexlab{a}})}\BibitemShut {NoStop}%
\bibitem [{\citenamefont {{Zhou}}\ \emph
  {et~al.}(2021{\natexlab{b}})\citenamefont {{Zhou}}, \citenamefont {{Xie}},
  \citenamefont {{Taniguchi}}, \citenamefont {{Watanabe}},\ and\ \citenamefont
  {{Young}}}]{Young2021II}%
  \BibitemOpen
  \bibfield  {author} {\bibinfo {author} {\bibfnamefont {H.}~\bibnamefont
  {{Zhou}}}, \bibinfo {author} {\bibfnamefont {T.}~\bibnamefont {{Xie}}},
  \bibinfo {author} {\bibfnamefont {T.}~\bibnamefont {{Taniguchi}}}, \bibinfo
  {author} {\bibfnamefont {K.}~\bibnamefont {{Watanabe}}},\ and\ \bibinfo
  {author} {\bibfnamefont {A.~F.}\ \bibnamefont {{Young}}},\ }\bibfield
  {title} {\bibinfo {title} {{Superconductivity in rhombohedral trilayer
  graphene}},\ }\href {https://doi.org/10.1038/s41586-021-03926-0} {\bibfield
  {journal} {\bibinfo  {journal} {\nat}\ }\textbf {\bibinfo {volume} {598}},\
  \bibinfo {pages} {434} (\bibinfo {year} {2021}{\natexlab{b}})}\BibitemShut
  {NoStop}%
\bibitem [{\citenamefont {{Zhou}}\ \emph {et~al.}(2022)\citenamefont {{Zhou}},
  \citenamefont {{Holleis}}, \citenamefont {{Saito}}, \citenamefont {{Cohen}},
  \citenamefont {{Huynh}}, \citenamefont {{Patterson}}, \citenamefont {{Yang}},
  \citenamefont {{Taniguchi}}, \citenamefont {{Watanabe}},\ and\ \citenamefont
  {{Young}}}]{Young2022}%
  \BibitemOpen
  \bibfield  {author} {\bibinfo {author} {\bibfnamefont {H.}~\bibnamefont
  {{Zhou}}}, \bibinfo {author} {\bibfnamefont {L.}~\bibnamefont {{Holleis}}},
  \bibinfo {author} {\bibfnamefont {Y.}~\bibnamefont {{Saito}}}, \bibinfo
  {author} {\bibfnamefont {L.}~\bibnamefont {{Cohen}}}, \bibinfo {author}
  {\bibfnamefont {W.}~\bibnamefont {{Huynh}}}, \bibinfo {author} {\bibfnamefont
  {C.~L.}\ \bibnamefont {{Patterson}}}, \bibinfo {author} {\bibfnamefont
  {F.}~\bibnamefont {{Yang}}}, \bibinfo {author} {\bibfnamefont
  {T.}~\bibnamefont {{Taniguchi}}}, \bibinfo {author} {\bibfnamefont
  {K.}~\bibnamefont {{Watanabe}}},\ and\ \bibinfo {author} {\bibfnamefont
  {A.~F.}\ \bibnamefont {{Young}}},\ }\bibfield  {title} {\bibinfo {title}
  {{Isospin magnetism and spin-polarized superconductivity in Bernal bilayer
  graphene}},\ }\href {https://doi.org/10.1126/science.abm8386} {\bibfield
  {journal} {\bibinfo  {journal} {Science}\ }\textbf {\bibinfo {volume}
  {375}},\ \bibinfo {pages} {774} (\bibinfo {year} {2022})}\BibitemShut
  {NoStop}%
\bibitem [{\citenamefont {{de la Barrera}}\ \emph {et~al.}(2022)\citenamefont
  {{de la Barrera}}, \citenamefont {{Aronson}}, \citenamefont {{Zheng}},
  \citenamefont {{Watanabe}}, \citenamefont {{Taniguchi}}, \citenamefont
  {{Ma}}, \citenamefont {{Jarillo-Herrero}},\ and\ \citenamefont
  {{Ashoori}}}]{Pablo2022}%
  \BibitemOpen
  \bibfield  {author} {\bibinfo {author} {\bibfnamefont {S.~C.}\ \bibnamefont
  {{de la Barrera}}}, \bibinfo {author} {\bibfnamefont {S.}~\bibnamefont
  {{Aronson}}}, \bibinfo {author} {\bibfnamefont {Z.}~\bibnamefont {{Zheng}}},
  \bibinfo {author} {\bibfnamefont {K.}~\bibnamefont {{Watanabe}}}, \bibinfo
  {author} {\bibfnamefont {T.}~\bibnamefont {{Taniguchi}}}, \bibinfo {author}
  {\bibfnamefont {Q.}~\bibnamefont {{Ma}}}, \bibinfo {author} {\bibfnamefont
  {P.}~\bibnamefont {{Jarillo-Herrero}}},\ and\ \bibinfo {author}
  {\bibfnamefont {R.}~\bibnamefont {{Ashoori}}},\ }\bibfield  {title} {\bibinfo
  {title} {{Cascade of isospin phase transitions in Bernal-stacked bilayer
  graphene at zero magnetic field}},\ }\href
  {https://doi.org/10.1038/s41567-022-01616-w} {\bibfield  {journal} {\bibinfo
  {journal} {Nat. Phys.}\ }\textbf {\bibinfo {volume} {18}},\ \bibinfo {pages}
  {771} (\bibinfo {year} {2022})}\BibitemShut {NoStop}%
\bibitem [{\citenamefont {{Seiler}}\ \emph {et~al.}(2022)\citenamefont
  {{Seiler}}, \citenamefont {{Geisenhof}}, \citenamefont {{Winterer}},
  \citenamefont {{Watanabe}}, \citenamefont {{Taniguchi}}, \citenamefont
  {{Xu}}, \citenamefont {{Zhang}},\ and\ \citenamefont {{Weitz}}}]{Weitz2022}%
  \BibitemOpen
  \bibfield  {author} {\bibinfo {author} {\bibfnamefont {A.~M.}\ \bibnamefont
  {{Seiler}}}, \bibinfo {author} {\bibfnamefont {F.~R.}\ \bibnamefont
  {{Geisenhof}}}, \bibinfo {author} {\bibfnamefont {F.}~\bibnamefont
  {{Winterer}}}, \bibinfo {author} {\bibfnamefont {K.}~\bibnamefont
  {{Watanabe}}}, \bibinfo {author} {\bibfnamefont {T.}~\bibnamefont
  {{Taniguchi}}}, \bibinfo {author} {\bibfnamefont {T.}~\bibnamefont {{Xu}}},
  \bibinfo {author} {\bibfnamefont {F.}~\bibnamefont {{Zhang}}},\ and\ \bibinfo
  {author} {\bibfnamefont {R.~T.}\ \bibnamefont {{Weitz}}},\ }\bibfield
  {title} {\bibinfo {title} {{Quantum cascade of correlated phases in
  trigonally warped bilayer graphene}},\ }\href
  {https://doi.org/10.1038/s41586-022-04937-1} {\bibfield  {journal} {\bibinfo
  {journal} {\nat}\ }\textbf {\bibinfo {volume} {608}},\ \bibinfo {pages} {298}
  (\bibinfo {year} {2022})}\BibitemShut {NoStop}%
\bibitem [{\citenamefont {Wehling}\ \emph {et~al.}(2011)\citenamefont
  {Wehling}, \citenamefont {\ifmmode \mbox{\c{S}}\else \c{S}\fi{}a\ifmmode
  \mbox{\c{s}}\else \c{s}\fi{}\ifmmode \imath \else \i
  \fi{}o\ifmmode~\breve{g}\else \u{g}\fi{}lu}, \citenamefont {Friedrich},
  \citenamefont {Lichtenstein}, \citenamefont {Katsnelson},\ and\ \citenamefont
  {Bl\"ugel}}]{Katsnelson2011PRL}%
  \BibitemOpen
  \bibfield  {author} {\bibinfo {author} {\bibfnamefont {T.~O.}\ \bibnamefont
  {Wehling}}, \bibinfo {author} {\bibfnamefont {E.}~\bibnamefont {\ifmmode
  \mbox{\c{S}}\else \c{S}\fi{}a\ifmmode \mbox{\c{s}}\else \c{s}\fi{}\ifmmode
  \imath \else \i \fi{}o\ifmmode~\breve{g}\else \u{g}\fi{}lu}}, \bibinfo
  {author} {\bibfnamefont {C.}~\bibnamefont {Friedrich}}, \bibinfo {author}
  {\bibfnamefont {A.~I.}\ \bibnamefont {Lichtenstein}}, \bibinfo {author}
  {\bibfnamefont {M.~I.}\ \bibnamefont {Katsnelson}},\ and\ \bibinfo {author}
  {\bibfnamefont {S.}~\bibnamefont {Bl\"ugel}},\ }\bibfield  {title} {\bibinfo
  {title} {{Strength of Effective Coulomb Interactions in Graphene and
  Graphite}},\ }\href {https://doi.org/10.1103/PhysRevLett.106.236805}
  {\bibfield  {journal} {\bibinfo  {journal} {Phys. Rev. Lett.}\ }\textbf
  {\bibinfo {volume} {106}},\ \bibinfo {pages} {236805} (\bibinfo {year}
  {2011})}\BibitemShut {NoStop}%
\bibitem [{\citenamefont {Roy}\ and\ \citenamefont {Yang}(2013)}]{RoyYang2013}%
  \BibitemOpen
  \bibfield  {author} {\bibinfo {author} {\bibfnamefont {B.}~\bibnamefont
  {Roy}}\ and\ \bibinfo {author} {\bibfnamefont {K.}~\bibnamefont {Yang}},\
  }\bibfield  {title} {\bibinfo {title} {{Bilayer graphene with parallel
  magnetic field and twisting: Phases and phase transitions in a highly tunable
  Dirac system}},\ }\href {https://doi.org/10.1103/PhysRevB.88.241107}
  {\bibfield  {journal} {\bibinfo  {journal} {Phys. Rev. B}\ }\textbf {\bibinfo
  {volume} {88}},\ \bibinfo {pages} {241107} (\bibinfo {year}
  {2013})}\BibitemShut {NoStop}%
\bibitem [{\citenamefont {Knox}\ \emph {et~al.}(2011)\citenamefont {Knox},
  \citenamefont {Locatelli}, \citenamefont {Yilmaz}, \citenamefont {Cvetko},
  \citenamefont {Mente\ifmmode~\mbox{\c{s}}\else \c{s}\fi{}}, \citenamefont
  {Ni\~no}, \citenamefont {Kim}, \citenamefont {Morgante},\ and\ \citenamefont
  {Osgood}}]{ARPESGraphene}%
  \BibitemOpen
  \bibfield  {author} {\bibinfo {author} {\bibfnamefont {K.~R.}\ \bibnamefont
  {Knox}}, \bibinfo {author} {\bibfnamefont {A.}~\bibnamefont {Locatelli}},
  \bibinfo {author} {\bibfnamefont {M.~B.}\ \bibnamefont {Yilmaz}}, \bibinfo
  {author} {\bibfnamefont {D.}~\bibnamefont {Cvetko}}, \bibinfo {author}
  {\bibfnamefont {T.~O.}\ \bibnamefont {Mente\ifmmode~\mbox{\c{s}}\else
  \c{s}\fi{}}}, \bibinfo {author} {\bibfnamefont {M.~A.}\ \bibnamefont
  {Ni\~no}}, \bibinfo {author} {\bibfnamefont {P.}~\bibnamefont {Kim}},
  \bibinfo {author} {\bibfnamefont {A.}~\bibnamefont {Morgante}},\ and\
  \bibinfo {author} {\bibfnamefont {R.~M.}\ \bibnamefont {Osgood}},\ }\bibfield
   {title} {\bibinfo {title} {{Making angle-resolved photoemission measurements
  on corrugated monolayer crystals: Suspended exfoliated single-crystal
  graphene}},\ }\href {https://doi.org/10.1103/PhysRevB.84.115401} {\bibfield
  {journal} {\bibinfo  {journal} {Phys. Rev. B}\ }\textbf {\bibinfo {volume}
  {84}},\ \bibinfo {pages} {115401} (\bibinfo {year} {2011})}\BibitemShut
  {NoStop}%
\bibitem [{\citenamefont {Brekke}\ \emph {et~al.}(2023)\citenamefont {Brekke},
  \citenamefont {Brataas},\ and\ \citenamefont {Sudb\o{}}}]{Brekke2023}%
  \BibitemOpen
  \bibfield  {author} {\bibinfo {author} {\bibfnamefont {B.}~\bibnamefont
  {Brekke}}, \bibinfo {author} {\bibfnamefont {A.}~\bibnamefont {Brataas}},\
  and\ \bibinfo {author} {\bibfnamefont {A.}~\bibnamefont {Sudb\o{}}},\
  }\bibfield  {title} {\bibinfo {title} {Two-dimensional altermagnets:
  Superconductivity in a minimal microscopic model},\ }\href
  {https://doi.org/10.1103/PhysRevB.108.224421} {\bibfield  {journal} {\bibinfo
   {journal} {Phys. Rev. B}\ }\textbf {\bibinfo {volume} {108}},\ \bibinfo
  {pages} {224421} (\bibinfo {year} {2023})}\BibitemShut {NoStop}%
\bibitem [{\citenamefont {Li}\ and\ \citenamefont {Liu}(2023)}]{Li2023}%
  \BibitemOpen
  \bibfield  {author} {\bibinfo {author} {\bibfnamefont {Y.-X.}\ \bibnamefont
  {Li}}\ and\ \bibinfo {author} {\bibfnamefont {C.-C.}\ \bibnamefont {Liu}},\
  }\bibfield  {title} {\bibinfo {title} {Majorana corner modes and tunable
  patterns in an altermagnet heterostructure},\ }\href
  {https://doi.org/10.1103/PhysRevB.108.205410} {\bibfield  {journal} {\bibinfo
   {journal} {Phys. Rev. B}\ }\textbf {\bibinfo {volume} {108}},\ \bibinfo
  {pages} {205410} (\bibinfo {year} {2023})}\BibitemShut {NoStop}%
\bibitem [{\citenamefont {Zhu}\ \emph {et~al.}(2023)\citenamefont {Zhu},
  \citenamefont {Zhuang}, \citenamefont {Wu},\ and\ \citenamefont
  {Yan}}]{Zhu2023}%
  \BibitemOpen
  \bibfield  {author} {\bibinfo {author} {\bibfnamefont {D.}~\bibnamefont
  {Zhu}}, \bibinfo {author} {\bibfnamefont {Z.-Y.}\ \bibnamefont {Zhuang}},
  \bibinfo {author} {\bibfnamefont {Z.}~\bibnamefont {Wu}},\ and\ \bibinfo
  {author} {\bibfnamefont {Z.}~\bibnamefont {Yan}},\ }\bibfield  {title}
  {\bibinfo {title} {{Topological superconductivity in two-dimensional
  altermagnetic metals}},\ }\href {https://doi.org/10.1103/PhysRevB.108.184505}
  {\bibfield  {journal} {\bibinfo  {journal} {Phys. Rev. B}\ }\textbf {\bibinfo
  {volume} {108}},\ \bibinfo {pages} {184505} (\bibinfo {year}
  {2023})}\BibitemShut {NoStop}%
\bibitem [{\citenamefont {Zhang}\ \emph
  {et~al.}(2024{\natexlab{b}})\citenamefont {Zhang}, \citenamefont {Hu},\ and\
  \citenamefont {Neupert}}]{Zhang2024}%
  \BibitemOpen
  \bibfield  {author} {\bibinfo {author} {\bibfnamefont {S.-B.}\ \bibnamefont
  {Zhang}}, \bibinfo {author} {\bibfnamefont {L.-H.}\ \bibnamefont {Hu}},\ and\
  \bibinfo {author} {\bibfnamefont {T.}~\bibnamefont {Neupert}},\ }\bibfield
  {title} {\bibinfo {title} {Finite-momentum cooper pairing in proximitized
  altermagnets},\ }\href {https://doi.org/10.1038/s41467-024-45951-3}
  {\bibfield  {journal} {\bibinfo  {journal} {Nat. Commun.}\ }\textbf {\bibinfo
  {volume} {15}},\ \bibinfo {pages} {1801} (\bibinfo {year}
  {2024}{\natexlab{b}})}\BibitemShut {NoStop}%
\bibitem [{\citenamefont {Ghorashi}\ \emph {et~al.}(2024)\citenamefont
  {Ghorashi}, \citenamefont {Hughes},\ and\ \citenamefont
  {Cano}}]{Ghorashi2023}%
  \BibitemOpen
  \bibfield  {author} {\bibinfo {author} {\bibfnamefont {S.~A.~A.}\
  \bibnamefont {Ghorashi}}, \bibinfo {author} {\bibfnamefont {T.~L.}\
  \bibnamefont {Hughes}},\ and\ \bibinfo {author} {\bibfnamefont
  {J.}~\bibnamefont {Cano}},\ }\bibfield  {title} {\bibinfo {title}
  {{Altermagnetic Routes to Majorana Modes in Zero Net Magnetization}},\ }\href
  {https://doi.org/10.1103/PhysRevLett.133.106601} {\bibfield  {journal}
  {\bibinfo  {journal} {Phys. Rev. Lett.}\ }\textbf {\bibinfo {volume} {133}},\
  \bibinfo {pages} {106601} (\bibinfo {year} {2024})}\BibitemShut {NoStop}%
\bibitem [{\citenamefont {Chakraborty}\ and\ \citenamefont
  {Black-Schaffer}(2024)}]{Chakraborty2023}%
  \BibitemOpen
  \bibfield  {author} {\bibinfo {author} {\bibfnamefont {D.}~\bibnamefont
  {Chakraborty}}\ and\ \bibinfo {author} {\bibfnamefont {A.~M.}\ \bibnamefont
  {Black-Schaffer}},\ }\bibfield  {title} {\bibinfo {title} {{Zero-field
  finite-momentum and field-induced superconductivity in altermagnets}},\
  }\href {https://doi.org/10.1103/PhysRevB.110.L060508} {\bibfield  {journal}
  {\bibinfo  {journal} {Phys. Rev. B}\ }\textbf {\bibinfo {volume} {110}},\
  \bibinfo {pages} {L060508} (\bibinfo {year} {2024})}\BibitemShut {NoStop}%
\bibitem [{\citenamefont {Kapitulnik}\ \emph {et~al.}(2009)\citenamefont
  {Kapitulnik}, \citenamefont {Xia}, \citenamefont {Schemm},\ and\
  \citenamefont {Palevski}}]{Kapitulnik2009}%
  \BibitemOpen
  \bibfield  {author} {\bibinfo {author} {\bibfnamefont {A.}~\bibnamefont
  {Kapitulnik}}, \bibinfo {author} {\bibfnamefont {J.}~\bibnamefont {Xia}},
  \bibinfo {author} {\bibfnamefont {E.}~\bibnamefont {Schemm}},\ and\ \bibinfo
  {author} {\bibfnamefont {A.}~\bibnamefont {Palevski}},\ }\bibfield  {title}
  {\bibinfo {title} {{Polar Kerr effect as probe for time-reversal symmetry
  breaking in unconventional superconductors}},\ }\href
  {https://doi.org/10.1088/1367-2630/11/5/055060} {\bibfield  {journal}
  {\bibinfo  {journal} {New J. Phys.}\ }\textbf {\bibinfo {volume} {11}},\
  \bibinfo {pages} {055060} (\bibinfo {year} {2009})}\BibitemShut {NoStop}%
\bibitem [{\citenamefont {Kruthoff}\ \emph {et~al.}(2017)\citenamefont
  {Kruthoff}, \citenamefont {de~Boer}, \citenamefont {van Wezel}, \citenamefont
  {Kane},\ and\ \citenamefont {Slager}}]{Slager2017}%
  \BibitemOpen
  \bibfield  {author} {\bibinfo {author} {\bibfnamefont {J.}~\bibnamefont
  {Kruthoff}}, \bibinfo {author} {\bibfnamefont {J.}~\bibnamefont {de~Boer}},
  \bibinfo {author} {\bibfnamefont {J.}~\bibnamefont {van Wezel}}, \bibinfo
  {author} {\bibfnamefont {C.~L.}\ \bibnamefont {Kane}},\ and\ \bibinfo
  {author} {\bibfnamefont {R.-J.}\ \bibnamefont {Slager}},\ }\bibfield  {title}
  {\bibinfo {title} {{Topological Classification of Crystalline Insulators
  through Band Structure Combinatorics}},\ }\href
  {https://doi.org/10.1103/PhysRevX.7.041069} {\bibfield  {journal} {\bibinfo
  {journal} {Phys. Rev. X}\ }\textbf {\bibinfo {volume} {7}},\ \bibinfo {pages}
  {041069} (\bibinfo {year} {2017})}\BibitemShut {NoStop}%
\bibitem [{\citenamefont {{Bradlyn}}\ \emph {et~al.}(2017)\citenamefont
  {{Bradlyn}}, \citenamefont {{Elcoro}}, \citenamefont {{Cano}}, \citenamefont
  {{Vergniory}}, \citenamefont {{Wang}}, \citenamefont {{Felser}},
  \citenamefont {{Aroyo}},\ and\ \citenamefont {{Bernevig}}}]{bernevig2017}%
  \BibitemOpen
  \bibfield  {author} {\bibinfo {author} {\bibfnamefont {B.}~\bibnamefont
  {{Bradlyn}}}, \bibinfo {author} {\bibfnamefont {L.}~\bibnamefont {{Elcoro}}},
  \bibinfo {author} {\bibfnamefont {J.}~\bibnamefont {{Cano}}}, \bibinfo
  {author} {\bibfnamefont {M.~G.}\ \bibnamefont {{Vergniory}}}, \bibinfo
  {author} {\bibfnamefont {Z.}~\bibnamefont {{Wang}}}, \bibinfo {author}
  {\bibfnamefont {C.}~\bibnamefont {{Felser}}}, \bibinfo {author}
  {\bibfnamefont {M.~I.}\ \bibnamefont {{Aroyo}}},\ and\ \bibinfo {author}
  {\bibfnamefont {B.~A.}\ \bibnamefont {{Bernevig}}},\ }\bibfield  {title}
  {\bibinfo {title} {{Topological quantum chemistry}},\ }\href
  {https://doi.org/10.1038/nature23268} {\bibfield  {journal} {\bibinfo
  {journal} {\nat}\ }\textbf {\bibinfo {volume} {547}},\ \bibinfo {pages} {298}
  (\bibinfo {year} {2017})}\BibitemShut {NoStop}%
\bibitem [{\citenamefont {{Po}}\ \emph {et~al.}(2017)\citenamefont {{Po}},
  \citenamefont {{Vishwanath}},\ and\ \citenamefont
  {{Watanabe}}}]{Vishwanath2017NatComm}%
  \BibitemOpen
  \bibfield  {author} {\bibinfo {author} {\bibfnamefont {H.~C.}\ \bibnamefont
  {{Po}}}, \bibinfo {author} {\bibfnamefont {A.}~\bibnamefont {{Vishwanath}}},\
  and\ \bibinfo {author} {\bibfnamefont {H.}~\bibnamefont {{Watanabe}}},\
  }\bibfield  {title} {\bibinfo {title} {{Complete theory of symmetry-based
  indicators of band topology}},\ }\href
  {https://doi.org/10.1038/s41467-017-00133-2} {\bibfield  {journal} {\bibinfo
  {journal} {Nat. Commun.}\ }\textbf {\bibinfo {volume} {8}},\ \bibinfo {eid}
  {50} (\bibinfo {year} {2017})}\BibitemShut {NoStop}%
\bibitem [{\citenamefont {Zhang}\ \emph {et~al.}(2019)\citenamefont {Zhang},
  \citenamefont {Jiang}, \citenamefont {Song}, \citenamefont {Huang},
  \citenamefont {He}, \citenamefont {Fang}, \citenamefont {Weng},\ and\
  \citenamefont {Fang}}]{Zhang2019}%
  \BibitemOpen
  \bibfield  {author} {\bibinfo {author} {\bibfnamefont {T.}~\bibnamefont
  {Zhang}}, \bibinfo {author} {\bibfnamefont {Y.}~\bibnamefont {Jiang}},
  \bibinfo {author} {\bibfnamefont {Z.}~\bibnamefont {Song}}, \bibinfo {author}
  {\bibfnamefont {H.}~\bibnamefont {Huang}}, \bibinfo {author} {\bibfnamefont
  {Y.}~\bibnamefont {He}}, \bibinfo {author} {\bibfnamefont {Z.}~\bibnamefont
  {Fang}}, \bibinfo {author} {\bibfnamefont {H.}~\bibnamefont {Weng}},\ and\
  \bibinfo {author} {\bibfnamefont {C.}~\bibnamefont {Fang}},\ }\bibfield
  {title} {\bibinfo {title} {{Catalogue of topological electronic materials}},\
  }\href {https://doi.org/10.1038/s41586-019-0944-6} {\bibfield  {journal}
  {\bibinfo  {journal} {\nat}\ }\textbf {\bibinfo {volume} {566}},\ \bibinfo
  {pages} {475} (\bibinfo {year} {2019})}\BibitemShut {NoStop}%
\bibitem [{\citenamefont {Vergniory}\ \emph {et~al.}(2019)\citenamefont
  {Vergniory}, \citenamefont {Elcoro}, \citenamefont {Felser}, \citenamefont
  {Regnault}, \citenamefont {Bernevig},\ and\ \citenamefont
  {Wang}}]{Vergniory2019}%
  \BibitemOpen
  \bibfield  {author} {\bibinfo {author} {\bibfnamefont {M.~G.}\ \bibnamefont
  {Vergniory}}, \bibinfo {author} {\bibfnamefont {L.}~\bibnamefont {Elcoro}},
  \bibinfo {author} {\bibfnamefont {C.}~\bibnamefont {Felser}}, \bibinfo
  {author} {\bibfnamefont {N.}~\bibnamefont {Regnault}}, \bibinfo {author}
  {\bibfnamefont {B.~A.}\ \bibnamefont {Bernevig}},\ and\ \bibinfo {author}
  {\bibfnamefont {Z.}~\bibnamefont {Wang}},\ }\bibfield  {title} {\bibinfo
  {title} {{A complete catalogue of high-quality topological materials}},\
  }\href {https://doi.org/10.1038/s41586-019-0954-4} {\bibfield  {journal}
  {\bibinfo  {journal} {\nat}\ }\textbf {\bibinfo {volume} {566}},\ \bibinfo
  {pages} {480} (\bibinfo {year} {2019})}\BibitemShut {NoStop}%
\bibitem [{\citenamefont {Tang}\ \emph {et~al.}(2019)\citenamefont {Tang},
  \citenamefont {Po}, \citenamefont {Vishwanath},\ and\ \citenamefont
  {Wan}}]{Tang2019}%
  \BibitemOpen
  \bibfield  {author} {\bibinfo {author} {\bibfnamefont {F.}~\bibnamefont
  {Tang}}, \bibinfo {author} {\bibfnamefont {H.~C.}\ \bibnamefont {Po}},
  \bibinfo {author} {\bibfnamefont {A.}~\bibnamefont {Vishwanath}},\ and\
  \bibinfo {author} {\bibfnamefont {X.}~\bibnamefont {Wan}},\ }\bibfield
  {title} {\bibinfo {title} {{Comprehensive search for topological materials
  using symmetry indicators}},\ }\href
  {https://doi.org/10.1038/s41586-019-0937-5} {\bibfield  {journal} {\bibinfo
  {journal} {\nat}\ }\textbf {\bibinfo {volume} {566}},\ \bibinfo {pages} {486}
  (\bibinfo {year} {2019})}\BibitemShut {NoStop}%
\bibitem [{\citenamefont {Mayorov}\ \emph {et~al.}(2011)\citenamefont
  {Mayorov}, \citenamefont {Elias}, \citenamefont {Mucha-Kruczynski},
  \citenamefont {Gorbachev}, \citenamefont {Tudorovskiy}, \citenamefont
  {Zhukov}, \citenamefont {Morozov}, \citenamefont {Katsnelson}, \citenamefont
  {Geim},\ and\ \citenamefont {Novoselov}}]{novoselov:nematic}%
  \BibitemOpen
  \bibfield  {author} {\bibinfo {author} {\bibfnamefont {A.~S.}\ \bibnamefont
  {Mayorov}}, \bibinfo {author} {\bibfnamefont {D.~C.}\ \bibnamefont {Elias}},
  \bibinfo {author} {\bibfnamefont {M.}~\bibnamefont {Mucha-Kruczynski}},
  \bibinfo {author} {\bibfnamefont {R.~V.}\ \bibnamefont {Gorbachev}}, \bibinfo
  {author} {\bibfnamefont {T.}~\bibnamefont {Tudorovskiy}}, \bibinfo {author}
  {\bibfnamefont {A.}~\bibnamefont {Zhukov}}, \bibinfo {author} {\bibfnamefont
  {S.~V.}\ \bibnamefont {Morozov}}, \bibinfo {author} {\bibfnamefont {M.~I.}\
  \bibnamefont {Katsnelson}}, \bibinfo {author} {\bibfnamefont {A.~K.}\
  \bibnamefont {Geim}},\ and\ \bibinfo {author} {\bibfnamefont {K.~S.}\
  \bibnamefont {Novoselov}},\ }\bibfield  {title} {\bibinfo {title}
  {{Interaction-Driven Spectrum Reconstruction in Bilayer Graphene}},\ }\href
  {https://doi.org/10.1126/science.1208683} {\bibfield  {journal} {\bibinfo
  {journal} {Science}\ }\textbf {\bibinfo {volume} {333}},\ \bibinfo {pages}
  {860} (\bibinfo {year} {2011})}\BibitemShut {NoStop}%
\bibitem [{\citenamefont {Weitz}\ \emph {et~al.}(2010)\citenamefont {Weitz},
  \citenamefont {Allen}, \citenamefont {Feldman}, \citenamefont {Martin},\ and\
  \citenamefont {Yacoby}}]{yacoby:nematic}%
  \BibitemOpen
  \bibfield  {author} {\bibinfo {author} {\bibfnamefont {R.~T.}\ \bibnamefont
  {Weitz}}, \bibinfo {author} {\bibfnamefont {M.~T.}\ \bibnamefont {Allen}},
  \bibinfo {author} {\bibfnamefont {B.~E.}\ \bibnamefont {Feldman}}, \bibinfo
  {author} {\bibfnamefont {J.}~\bibnamefont {Martin}},\ and\ \bibinfo {author}
  {\bibfnamefont {A.}~\bibnamefont {Yacoby}},\ }\bibfield  {title} {\bibinfo
  {title} {{Broken-Symmetry States in Doubly Gated Suspended Bilayer
  Graphene}},\ }\href {https://doi.org/10.1126/science.1194988} {\bibfield
  {journal} {\bibinfo  {journal} {Science}\ }\textbf {\bibinfo {volume}
  {330}},\ \bibinfo {pages} {812} (\bibinfo {year} {2010})}\BibitemShut
  {NoStop}%
\end{thebibliography}%

\end{document}